\documentclass{ar-1col}

\usepackage[comma]{natbib}
\usepackage{amsfonts}
\usepackage{amsmath}
\usepackage{amssymb, graphicx}

\newcommand{\bDiamond}{\mathbin{\Diamond}}
\newcommand{\bLozenge}{\mathbin{\blacklozenge}}

\usepackage{tikz}
\usepackage{pgfplots}
\usepackage{forest}
\usetikzlibrary{fit,positioning}

\definecolor{headcolor}{cmyk}{0,1.0,1.0,0.30}

\usepackage{tikz-qtree,tikz-qtree-compat}
\usetikzlibrary{calc}

\usepackage{url}
\usepackage{todonotes}

\usepackage{natbib}
\jname{Xxxx. Xxx. Xxx. Xxx.}
\jvol{AA}
\jyear{YYYY}
\doi{10.1146/((please add article doi))}

\newcommand{\bv}{\mathbf{v}}
\newcommand{\bb}{\mathbf{b}}

\newcommand{\bF}{\mathbf{F}}
\newcommand{\bI}{\mathbf{I}}
\newcommand{\bP}{\mathbf{P}}
\newcommand{\bone}{\mbox{\boldmath $1$}}

\usepackage{soul}

\begin{document}

\markboth{Salerno and Li}{High-Dimensional Survival Analysis}

\title{High-Dimensional Survival Analysis: Methods and Applications}

\author{Stephen Salerno$^1$ and Yi Li$^1$
\affil{$^1$Department of Biostatistics, University of Michigan, Ann Arbor, United States, 48109; email: yili@umich.edu}}

\begin{abstract}
In the era of precision medicine, time-to-event outcomes such as time to death or progression are routinely collected, along with high-throughput covariates. These high-dimensional data defy classical survival regression models, which are either infeasible to fit or likely to incur low predictability due to over-fitting. To overcome this, recent emphasis has been placed on developing novel approaches for feature selection and survival prognostication. We will review various cutting-edge methods that handle survival outcome data with high-dimensional predictors, highlighting recent innovations in machine learning approaches for survival prediction. We will cover the statistical intuitions and principles behind these methods and conclude with extensions to more complex settings, where competing events are observed. We exemplify these methods with applications to the Boston Lung Cancer Survival Cohort study, one of the largest cancer epidemiology cohorts investigating the complex mechanisms of lung cancer.
\end{abstract}


\maketitle



\section{INTRODUCTION}

Survival analysis is an area of statistics where the random variate is {\it survival time} or the time until the occurrence of a specific event, which represents a qualitative change or the transition from one discrete state to another (e.g., alive to deceased). The most often studied event in biomedicine is death, though events of interest in fields ranging from sociology to industry, to engineering, to finance, to astronomy are widely encountered. The goals of survival analysis are to describe the probability of an event occurring by some time, to detect associations between risk factors and events, or to predict survival times based on informative characteristics. What distinguishes survival outcomes from other outcomes is the presence of {\it censoring}, meaning that the event of interest may not be observed for all subjects;  subjects whose event times are not observed are said to be {\it censored}. In practice, the fraction of event times that are censored in a study population can be substantial, prohibiting the direct use of standard regression methods. Estimation methods in survival analysis are built around extracting information from all subjects, censored or not.

In the era of precision medicine, survival outcomes with high-throughput covariates or predictors are routinely collected. These {\it high-dimensional data} (i.e., with the number of predictors exceeding the number of observations) challenge classical survival regression models, which are either infeasible to fit or likely to incur low predictability due to over-fitting. Recent emphasis has been placed on developing novel approaches for feature selection and survival prognostication. We will review various methods that handle survival outcome data with high-dimensional predictors, highlighting recently developed machine learning approaches for survival prediction. We will also discuss recent developments for deep learning in survival settings and introduce some new deep learning techniques in the presence of competing or semi-competing outcomes. A competing risk is an event whose occurrence precludes the occurrence of another event of interest \citep{austin2017practical}, while in a semi-competing setting, the occurrence of a non-terminal event (e.g., disease progression) is subject to a terminal event (e.g., death), but not vice versa \citep{haneuse2016semi}. We illustrate a novel deep learning approach for prediction under semi-competing outcomes, and  exemplify the method using data from the  Boston Lung Cancer Survival Cohort (BLCSC), a large hospital-based cancer epidemiology cohort investigating the molecular mechanisms and clinical pathophysiology of lung cancer \citep{Christiani2017blcs}.
 
This paper is outlined as follows. In Section 2, we provide a brief overview of some key concepts and notation in survival analysis and introduce the necessary perquisites on which much of the subsequent literature is built. In Section 3, we survey current techniques for fitting survival models with high-dimensional covariates, primarily focusing on methods that perform feature selection under sparsity assumptions. We briefly discuss ultra high-dimensional settings and introduce screening methods, and end this section with a discussion of methods for drawing valid inference with high-dimensional covariates. In Section 4, we turn to machine learning for survival prediction. We first discuss the application of common machine learning concepts in these settings, such as a support vector machines, recursive partitioning and survival trees, and ensemble learners such as random survival forests. We briefly review artificial neural networks and extend this notion to survival prediction. In Section 5, we review existing deep learning procedures for competing risk analysis, illustrate a new deep learning approach for predicting semi-competing outcomes, and work through the BLCSC study. We conclude with remarks on future work and open areas. The online Supplementary Material tabulates the reviewed methods and their available software, and presents additional simulation results. 


\section{NOTATION}

Consider a study consisting of $n$ subjects. The outcome variable is the time to the event of interest, such as death or cancer progression. Events in other contexts can be bankruptcy, COVID-19 infection, graduation, missing a mortgage payment, etc. A {\it time zero} also needs to be set carefully, to have proper biological or practical interpretations when helping to address specific scientific questions. For instance, some common choices of time zero in medical studies include date of birth, time of diagnosis, date of randomization in a clinical trial, or first date receiving a treatment. A unique aspect of survival analysis is that the event may go unobserved for some individuals. In particular, {\it right censoring} occurs when a subject's follow-up ends before the event can be observed ({\bf Figure \ref{fig:1}}). Though other types of censoring exist, we focus on right censoring, which happens most often in practice. 

We denote the $i$th subject's survival and censoring times respectively by $T_i$ and $C_i$ $(i = 1,\ldots, n)$, which are non-negative random variates.~For the $i$th subject, we observe $\boldsymbol{X}_i$, a $p$-vector of covariates, $Y_i = \min(T_i, C_i)$, and the event indicator $\delta_i = \mathbb{I}(T_i \leq C_i)$, where $\mathbb{I}(\cdot)$ is an indicator function.~We assume that subjects are independent from each other, and that $T_i \perp C_i$, given $\boldsymbol{X}_i$. Often, the goal of survival analysis is to  associate $\boldsymbol{X}_i$  with the distribution of $T_i$, and, in particular, model the conditional hazard function given $\boldsymbol{X}_i$, i.e., 
\begin{equation} \label{condhaz}
 \lambda(t | \boldsymbol{X}_i) = \lim _{\Delta  \rightarrow 0} \frac{1}{\Delta} \operatorname{Pr}( t \le T_i  < t + \Delta  | T_i
 \ge t, \boldsymbol{X}_i),
\end{equation}
which measures the instantaneous failure rate at a given time among those who are alive and with $\boldsymbol{X}_i$. Throughout this review, for simplicity, we assume that $\boldsymbol{X}_i$ is time-invariant, though in many circumstances extensions to time-dependent covariates  are possible. 

\begin{figure}[!ht]
    \centering
    \begin{tikzpicture}[x = 1cm, y = 0.4cm]
        \draw[-latex] (0,0)--(9,0) node [right] {Time}; 
        \draw[-]      (1,0)--(1,5) node [above] {}; 
        \draw[-]      (8,0)--(8,5) node [above] {}; 
        \draw[|-]     (2,4)--(5,4) node [right] {\hskip 2ex Death}; 
        \draw[|-]     (4,2)--(8,2) node [right] {\hskip 2ex Censoring};
        \node[]          at (1,-1) {Study Start};
        \node[]          at (8,-1) {Study End};
        \node[]          at (0, 4) {Patient 1};
        \node[headcolor] at (5, 4) {\huge $\times$};
        \node[]          at (0, 2) {Patient 2};
        \node[white]     at (8, 2) {\huge \textbullet};
        \node[]          at (8, 2) {\huge $\circ$};
    \end{tikzpicture}
    \vspace{2ex}
    \caption{Schematic of observations for two example patients, with different entry times, over the course of a study. The event of interest, death, is observed for Patient 1, whereas Patient 2 is censored, as the patient is still alive at the end of the study.}
    \label{fig:1}
\end{figure}


\section{HIGH-DIMENSIONAL SURVIVAL MODELS}

In high-dimensional settings, it is not recommended to build prediction models with all of the available features due to the risk of {\it over-fitting}. A useful strategy is to select only the most vital features under the assumption of {\it sparsity}, meaning that most of the potential predictors are `unimportant,' with nearly no effect on the outcome \citep{friedman2010regularization}. A key question is how to perform variable selection and estimation simultaneously, and the most widely used approaches fall under the class of {\it regularized} regression models. Regularization refers to the addition of a penalty term to the objective function, which shrinks the coefficient estimates toward zero and possibly forces some of them to be exactly zero. This mitigates over-fitting and results in parsimonious prediction models \citep{Tibshirani1996}.

\subsection{Regularized Cox Models} 

The approach that dominates survival analysis in the biomedical literature is the \cite{Cox1972} proportional hazards model, famed for presenting both a novel hazard model and a novel concept of {\it partial likelihood}. The model links the conditional hazard function (\ref{condhaz}) to $\boldsymbol{X}_i$ via
\[
    \lambda(t | \boldsymbol{X}_i) = \lambda_0(t) \exp (\boldsymbol{X}_i^\top \boldsymbol{\beta}),
\]
where the baseline hazard, $\lambda_0(t)$, is unspecified, and $\boldsymbol{\beta}
= (\beta_1, \ldots, \beta_p)^\top$ is the coefficient vector of $\boldsymbol{X}_i$ to be estimated, with a fixed $p<n$, by maximizing the partial likelihood, i.e.,
\[
    PL(\boldsymbol{\beta}) = \prod_{i: \delta_i=1 } PL_i(\boldsymbol{\beta}),
\]
 with $PL_i(\boldsymbol{\beta})$ being the contribution for subject $i$ who is observed to die:
\[
    PL_i(\boldsymbol{\beta}) = \operatorname{Pr}(\mbox{subject $i$ dies at
    $Y_i$}\ |\ \mbox{someone  from } {\cal R}(Y_i)  \mbox{ dies at $Y_i$} ) = \frac{\exp (\boldsymbol{X}_i^\top \boldsymbol{\beta})}{\sum_{j\in {\cal
    R}(Y_i)}\exp(\boldsymbol{X}_j^\top \boldsymbol{\beta})}
\]
where ${\cal R}(Y_i) = \{ j: Y_j \ge Y_i$\}. In high-dimensional settings, that is, $p>n$ and asymptotically $p$ and $n$ may both go to infinity \citep{zhao2006model}, directly optimizing the partial likelihood is not feasible because of over-parameterization. Instead, regularized regression adds a penalty term to the negative log partial likelihood, $\ell(\boldsymbol{\beta})$, and optimizes a penalized version of objective function:
\[
    - \ell(\boldsymbol{\beta}) + \eta Pen(\boldsymbol{\beta}),
\]
where the penalty $Pen(\boldsymbol{\beta})$ is controlled by a positive tuning parameter, $\eta$, to be selected through cross-validation. A widely recognized family of penalties is based on the $l_q$-norm,
\[
    ||\boldsymbol{\beta}||_q = ( \sum_{j=1}^p |\beta_j|^q)^{1/q},\ q\geq 0.
\]

Regularization approaches with $Pen(\boldsymbol{\beta})=||\boldsymbol{\beta}||_2^2$, known as ridge regression \citep{hoerl1970ridge}, are applied to the Cox model, and return unique and shrunk estimates \citep{verweij1994penalized}. However, ridge regression does not promote sparsity, as it cannot shrink individual coefficients to zero. The least absolute shrinkage and selection operator (LASSO) \citep{Tibshirani1996}, with $Pen(\boldsymbol{\beta})=||\boldsymbol{\beta}||_1$, penalizes the absolute sum of the coefficient estimates and has been routinely used for producing sparse models. Its application \citep{Tibshirani1997} to  survival settings, namely, Cox LASSO, has become a widely used approach for high-dimensional survival analysis by performing feature selection and estimation simultaneously ({\bf Figure \ref{fig:2}}).

\begin{figure}[!ht]
    \centering
    \begin{tikzpicture} 
        \draw[-latex] (-2,0) -- (2,0) node[below]{$\beta_1$};
        \draw[-latex] (0,-1.5) -- (0,2) node[left]{$\beta_2$};
        \draw[thick,cyan] (-1,0) -- (0,-1) -- (1,0) -- (0,1) -- cycle;
        \coordinate[] (beta1) at (1.75,2);
        \draw node[] (lab1) at (2.5,1.5){$\hat{\beta}_{MPLE}$};
        \foreach \X in {0.5,0.75,1.0}
        {\draw[rotate=30,headcolor!70] (beta1) circle({2*\X} and {0.5*\X});}
        \fill (beta1) circle (1mm);
        \coordinate[] (beta2) at (0,1);
        \draw node[] (lab2) at (0.75,0.6){$\hat{\beta}_{LASSO}$};
        \fill (beta2) circle (1mm);
    \end{tikzpicture}
    \caption{Graphical representation of Cox LASSO with two dimensional predictors. The blue diamond represents the constraint region $|\beta_1| + |\beta_2| \leq s$ for a given $s$. $\hat{\beta}_{MPLE}$ and $\hat{\beta}_{LASSO}$  represent the maximum partial likelihood and Cox LASSO estimates, respectively, and the red ellipses are contours of the partial likelihood function. As shown, subject to the $l_1$ constraint,  $\hat{\beta}_{LASSO}$ is shrunk to zero compared to  $\hat{\beta}_{MPLE}$, and  Cox LASSO estimates  $\beta_1$  to be exactly zero.}
    \label{fig:2}
\end{figure}

LASSO has several notable statistical properties. It possesses model selection consistency under certain regularity conditions, in particular, the strong irrepresentable
condition when $p$  grows much faster than $n$ (i.e., that the absolute sum of coefficients for the regression of any noise variable on signal variables must be strictly smaller than 1) \citep{zhao2006model}. It has a Bayesian interpretation by viewing $\boldsymbol{\beta}$ as having a ``double exponential" prior \citep{Tibshirani2009}. However, as the LASSO penalty term is linear in the size of the coefficients, it leads to biased estimates, especially for the coefficients with large absolute values. To remedy this, \cite{ZhangLu2007} proposed the adaptive Cox LASSO  by utilizing $Pen(\boldsymbol{\beta}) = \sum_j w_j |\beta_j|$, with smaller weights, $w_j$, assigned to larger coefficients and vice versa. The estimates are $\sqrt{n}$ consistent if $\sqrt{n}\eta = O(1)$ and have oracle properties if $\sqrt{n}\eta \rightarrow 0$ and $n\eta \rightarrow \infty$. When $p>n$, it was suggested to use robust estimates such as ridge regression estimates to determine the $w_j$'s.

\cite{FanLi2002} proposed a smoothly clipped absolute deviation (SCAD) penalty, which is a quadratic spline function of $|\beta|$ with knots at $\eta$ and $\alpha\eta$. Its derivative w.r.t $|\beta|$, i.e.,
\[
    \eta\left\{\mathbb{I}(|\beta| \leq \eta) + \frac{(\alpha\eta-|\beta|)_+}{(\alpha-1)\eta} \mathbb{I}(|\beta| > \eta)\right\};\quad \alpha > 2, \eta>0, 
\]
may more clearly show the role of the penalty in regularizing estimating equations \citep{FanLi2002}. While the SCAD penalty retains the penalization rate of LASSO for small coefficients, it relaxes the rate of penalization smoothly as the absolute value of the coefficient increases. Asymptotically, the SCAD penalty yields $\sqrt{n}$-consistent  estimates (with a proper rate of $\eta$) and possesses oracle properties (if $\sqrt{n}\eta \rightarrow 0$ and $n\eta \rightarrow \infty$). Strong oracle properties for LASSO and SCAD were established in \cite{bradic2011regularization}, which further proposed a class of nonconvex penalization procedures for the Cox model. Nonconvex regularization, including SCAD, is appealing as it obtains support recovery properties under much weaker assumptions than for $l_1$ penalization \citep{loh2017support}.

Another extension is the elastic net penalty for Cox models \citep{wu2012elastic}, which combines the LASSO and ridge penalties; but, unlike LASSO, is capable of selecting more predictors than the sample size \citep{zou2005regularization}. This notion was generalized by \cite{vinzamuri2013cox} with the kernel elastic net Cox regression model,  which replaces the ridge penalty with $\boldsymbol{\beta}^\top \boldsymbol{\Sigma} \boldsymbol{\beta}$. Here, $\boldsymbol{\Sigma}$ is a $p\times p$ radial basis function kernel matrix of predictors, which measures pairwise similarity between predictors. This penalty is meant to encourage correlated predictors to have similar strengths on survival prediction. 
 Other regularized Cox methods include the group Cox LASSO \citep{kim2012analysis}, which selects groups of related covariates as a whole, and the fused LASSO \citep{chaturvedi2014fused}, which penalizes both the coefficient estimates and their successive differences for ordered features ({\bf Table \ref{tab:1}}).

\begin{table}[!ht]
    \tabcolsep7.5pt
    \caption{Examples of regularized Cox regression methods and their penalty terms.}
    \label{tab:1}
    \begin{center}
    \begin{tabular}{l|c|c}
        \hline
        Method & Penalty & Notes\\
        \hline
        Ridge & $||\boldsymbol{\beta}||_2^2$ & -\\
        LASSO & $||\boldsymbol{\beta}||_1$   & -\\
        Elastic Net & $\alpha ||\boldsymbol{\beta}||_1 + (1-\alpha)||\boldsymbol{\beta}||_2^2$ & $0 < \alpha < 1$\\
        Adaptive LASSO & $\sum_j w_j |\beta_j|$ & $w_j \geq 0$\\
        SCAD ($Pen(|\beta|)$) & $ Pen'(|\beta|) =\eta\left\{\mathbb{I}(|\beta| \leq \eta) + \frac{(\alpha\eta-|\beta|)_+}{(\alpha-1)\eta} \mathbb{I}(|\beta| > \eta)\right\}$ & $\alpha > 2, \eta>0$\\
        Group LASSO & $\sum_g ||\boldsymbol{\beta}_g||_1$ &  
        $\boldsymbol{\beta}_g = (\beta_{g1}, \ldots, \beta_{gj_g})^\top$\\
        Fused LASSO & $\sum_j |\beta_j|$ and $\sum_j |\beta_j - \beta_{j-1}|$ & -\\
        \hline
    \end{tabular}
    \end{center}
\end{table}

\subsection{The Dantzig Selector for Survival Data}

\noindent \cite{candes2007dantzig} proposed another type of regularized estimator known as the Dantzig selector for linear regression:
\begin{equation} \label{linreg}
    \boldsymbol{Y} = \boldsymbol{X}\boldsymbol{\beta} + \boldsymbol{\epsilon},
\end{equation}
where $\boldsymbol{Y}$, $\boldsymbol{X}$, $\boldsymbol{\beta}$, and $\boldsymbol{\epsilon}$ are an $n \times 1$ vector of responses, an $n \times p$ covariate matrix, a $p \times 1$ vector of coefficients and an $n \times 1$ vector of zero-mean residual errors, respectively. It estimates $\boldsymbol{\beta}$ by solving
\[
\begin{gathered}
    \min ||\boldsymbol{\beta}||_1 \\
    \vspace{2ex}
    \text{subject to } ||\boldsymbol{X}^\top (\boldsymbol{Y}-\boldsymbol{X}\boldsymbol{\beta})||_{\infty} \le \eta_Q,\\
\end{gathered}
\]
where $\eta_Q>0$ is a tuning parameter.~Empowered by linear programming, the Dantzig selector offers a useful alternative as a regularized estimating equation approach. As a dual problem of LASSO, it often produces the same solution path \citep{candes2007dantzig}.

On the other hand, accelerated failure time (AFT) models have become a useful alternative to Cox models due to the ease of interpretation \citep{saikia2017review}. An AFT model links the (log transformed) survival time to covariates via a linear model 
\begin{equation} \label{aftmodel}
   \log(T_i) = \boldsymbol{X}_i^\top \boldsymbol{\beta} + e_i, 
\end{equation}
where the $\log$ transformation ensures the parameter space of $\boldsymbol{\beta}$ is unconstrained, and the distribution of the errors, $e_i$, induces a distribution for $T_i$ ({\bf Table \ref{tab:2}}).~For parametric AFT models, maximum likelihood estimation (MLE) can be used for inference. When $e_i$'s distribution is unspecified, the models are semi-parametric and the MLE is difficult to obtain, as the likelihood involves infinite dimensional parameters. With a fixed $p < n$, \cite{buckley1979linear} proposed an estimating equation approach by imputing the censored outcomes, and solving a least squares problem.

\begin{table}[!ht]
    \centering
    \caption{Specifications of various parametric accelerated failure time models.}
    \vspace{2ex}
    \begin{tabular}{c|c}
    \hline
    Distribution of $e_i$ & Induced Distribution of $T_i$ \\
    \hline
    Normal Distribution & Log-Normal Distribution \\
    Extreme Value Distribution & Weibull Distribution \\
    Logistic Distribution & Log-Logistic Distribution \\
    \hline
    \end{tabular}
    \label{tab:2}
\end{table}

For AFT models with high-dimensional predictors, one cannot directly apply LASSO estimation, as the objective function again involves infinite dimensional parameters. Motivated by the work of \cite{candes2007dantzig} for regularized least squares estimation, \cite{li2014dantzig} used Buckley-James imputation to express AFT estimation as a least squares problem and then applied the Dantzig selector: 
\[
\begin{gathered}
    \min ||\boldsymbol{\beta}||_1 \\
    \vspace{2ex}
    \text{subject to } ||\boldsymbol{X}^\top \bP_n \{\boldsymbol{T}^*(\boldsymbol{\beta})-\boldsymbol{X}\boldsymbol{\beta}\}||_{\infty} \le \eta_Q,\\
\end{gathered}
\]
where $T_i^*(\boldsymbol{\beta}) = \log(Y_i) + (1-\delta_i)\frac{\int_{e_i(\boldsymbol{\beta})}^\infty \hat{S}(s, \boldsymbol{\beta})ds}{\hat{S}(e_i(\boldsymbol{\beta}), \boldsymbol{\beta})}$ are imputed outcomes with the Kaplan-Meier estimate, $\hat{S}(\cdot, \boldsymbol{\beta})$,  based on $\{e_i(\boldsymbol{\beta}) = \log(Y_i) - \boldsymbol{\beta}^\top\boldsymbol{X}_i, \delta_i\}, i=1, \ldots, n$, and $\boldsymbol{T}^*(\boldsymbol{\beta}) = (T_1^*(\boldsymbol{\beta}), \ldots, T_n^*(\boldsymbol{\beta}))^\top$.  Here, the projection matrix $\bP_n= \bI_n - \bone \bone^\top/n$, where $\bI_n$ and $\bone$ are an $n \times n$ identity matrix and
an $n \times 1$ vector of 1's, 
is for centering covariates to avoid estimation of the intercept or the expection of $e_i$ in model (\ref{aftmodel}). An iterative approach is necessary because this is not a linear programming problem, and, like the Dantzig selector for linear models, estimates may be biased and may not possess the oracle property. 

To address this, \cite{li2014dantzig} considered an adaptive version of Dantzig selector with data-driven weights that vary inversely with the magnitude of coefficients. They showed that the weighted Dantzig selector has model selection consistency and oracle properties. On the other hand, a Dantzig selector for the Cox model was proposed by \cite{antoniadis2010dantzig} based on partial likelihood score equations.

Note that, in {\it ultra} high-dimensional settings where $p \gg n$, penalized variable selection methods such as those described so far may incur high computational costs, numeric instability, and poor reproducibility \citep{fan2008sure}. As such, variable screening is a crucial first step in identifying predictive biomarkers and reducing the dimensionality of the feature space before applying regularized methods. Feature screening methods such as sure independence screen (SIS) \citep{fan2008sure} fit marginal regression models for each covariate one at a time, choose a threshold, and retain those covariates with magnitudes of marginal effects above the threshold. In the ultra high-dimensional survival settings, additional censoring issues need to be addressed. Recent advancements in survival feature screening have included sure independence screening \citep{fan2010high}, principled sure independent screening \citep{zhao2012principled}, score test screening \citep{zhao2014score}, concordance-measure based screening \citep{ma2017concordance}, Buckley-James assisted sure screening \citep{liu2020new}, conditional screening \citep{kang2017partition,hong2018conditional},  integrated power density screening \citep{hong2018integrated}, $L_q$ norm screening \citep{hong2020lq}, and forward regression \citep{hong2019quantile, pijyan2020consistent}. See a focused review of survival feature screening in \cite{hong2017feature}.

\subsection{Inference with High-Dimensional Covariates}

As simultaneous estimation and inference is challenging within the high-dimensional survival framework, we review limited methods available for drawing inference in this area. More broadly, high-dimensional regression inference methods largely fall under post-selection inference and debiased LASSO estimation. Further challenges arise in that post-selection inference is conditional on the selected subset and does not account for variation in model selection. Several authors used debiased LASSO  \citep{van2014asymptotically, javanmard2014confidence, yu2018confidence} to draw inference; however, these methods require estimation of the inverse of a $p \times p$  information matrix, which is a daunting task, especially when $p>n$ \citep{xia2021debiased, xia2021statistical}. 

\subsubsection{Selection-Assisted Partial Regression and Smoothing (SPARES)} 

To address this challenge, \cite{fei2019drawing} proposed SPARES to draw inference for high-dimensional linear models (\ref{linreg}) with $p>n$. Under this framework, model selection and partial regression are conducted separately on partitioned data, and multiple sample splittings or {\it boostraps} are used to account for variations in variable selection and estimation. Specifically, given data ${\cal D} = (\boldsymbol{X}, \boldsymbol{Y})$ and a variable selection procedure $\mathcal{S}_\eta$, data are split into equally sized ${\cal D}_1$ and ${\cal D}_2$. Denote the variables selected by $\mathcal{S}_\eta$ on ${\cal D}_2$ as $S = \mathcal{S}_\eta({\cal D}_2)$. On ${\cal D}_1 = (\boldsymbol{X}^1, \boldsymbol{Y}^1)$ and for any $j \in \{1, \ldots, p\}$, $\boldsymbol{Y}^1$ is regressed on $\boldsymbol{X}^1_{S\cup j}$ to estimate $\beta^0_j$  by
\begin{equation}\label{spare}
    \tilde{\beta}_j = \Big\{ ({\boldsymbol{X}^1_{S\cup j}}^\top\boldsymbol{X}^1_{S\cup j})^{-1}{\boldsymbol{X}^1_{S\cup j}}^\top \boldsymbol{Y}^1 \Big\}_{j},
\end{equation}	
 where $\{\cdot\}_j$ denotes the estimate corresponding to variable $j$. Equation (\ref{spare})
 is termed the {\it partial regression estimator} \citep{fei2019drawing}.
 Set $\hat{\boldsymbol{\beta}} = (\tilde{\beta}_1, \ldots,  \tilde{\beta}_p)^\top$.  The rationale behind this idea is that, if $\operatorname{Pr}(S\supset S_{0}) \rightarrow 1$, where $S_0$ is the true active set, the ``one-time" partial regression in (\ref{spare}) returns a consistent estimator for $\beta^0_j$, regardless of whether $j\in S$. However, the one-time estimator is highly variable, heavily depending on ${S}$ and the specific data-split, and does not account for variation in the variables selected. To address this,  data are randomly split, say,  $B$ times, and partial regression are carried out over these $B$ random splits.
  Denote by  $\hat{\boldsymbol{\beta}}^b$ the estimate of $\boldsymbol{\beta}$ based on the $b$th resample ($b=1, \ldots, B$). The SPARES estimator is
\[
    \hat{\boldsymbol{\beta}} = \frac{1}{B} \sum_{b=1}^{B}\hat{\boldsymbol{\beta}}^b.
\]
To draw inference, a nonparametric delta method \citep{efron2014estimation,van2000asymptotic} is used to estimate the standard error of $\hat{\beta}_j\;(j=1,\ldots,p)$ as $\hat{\mathrm{se}}^B_j=[\sum_{i=1}^{n}\hat{\mathrm{cov}}^2_{ij}]^{1/2}$, where $ \hat{\mathrm{cov}}_{ij}$ is the sample covariance between $I_{bi}$ and $\hat{\beta}^b_j$, with  $I_{bi}$ indicating whether subject $i$ is included in the $b$th resample (used for partial regression). Approximate $95\%$ confidence intervals are given by $\hat{\beta}_j \pm 1.96\ \hat{\mathrm{se}}^B_j$, while a two-sided $p$-value testing $H_0: \beta_j = 0$ is given by $2\times [1-\Phi(|\hat{\beta}_j|/\hat{\mathrm{se}}^B_j)]$, where $\Phi (\cdot)$ is the distribution function of $N(0,1)$. SPARES provides a novel inference technique that converts a high-dimensional problem to a low-dimensional regression. It is valid with general selection methods, including  LASSO, SCAD, screening, and boosting, as long as they possess selection consistency or the relaxed ``sure screening" condition in \cite{fei2021estimation}. Further, this approach is not sensitive to the tuning parameter $\eta$ in $\mathcal{S}_\eta$ and can be extended to analyze censored outcomes, as detailed below.

\subsubsection{High-Dimensional Censored Quantile Regression}
As opposed to the Cox and AFT models, censored quantile regression permits the effects of covariates to vary across quantile levels, thus accommodating potentially heterogeneous impacts of certain risk predictors. For $\tau \in (0, 1)$, the $\tau$-th quantile is a value at or below which a $\tau$-fraction of population lies. Denote the $\tau$-th conditional quantile of $ \tilde{T} =\log(T)$ given $\boldsymbol{X}_i$ by $Q_{\tilde{T}}(\tau|\boldsymbol{X}_i)$. The censored quantile regression model \citep{powell1986censored, portnoy2003censored} stipulates that
\begin{equation} \label{cqr}
    Q_{\tilde{T}}(\tau|\boldsymbol{X}_i) = \beta_0(\tau) + \boldsymbol{X}_i^\top\boldsymbol{\beta}(\tau),
\end{equation}
where $\boldsymbol{\beta}(\tau)$ is a vector of quantile-specific regression coefficients.
With a fixed $p<n$,  
\cite{peng2008survival} proposed a class of martingale-based estimating equations
to estimate $\boldsymbol{\beta}(\tau)$ in  model (\ref{cqr}) over a fine grid of quantile levels, that is, $\Gamma_m=\{\nu=\tau_0, \tau_1, \ldots , \tau_m=\tau_U\}$, where $0<\tau_U<1$ is an upper bound for estimability.
Specifically, $\boldsymbol{\beta}(\tau_k)$'s, where $\tau_k \in \Gamma_m,$  can be sequentially and consistently estimated by solving
\begin{equation} \label{cqrlow}
\sum_{i=1}^{n}\boldsymbol{X}_i\left(N_i\left(\boldsymbol{X}_{i}^\top\boldsymbol{\beta}(\tau_k) \right) - \sum_{r= 0}^{k-1} \int_{\tau_r}^{\tau_{r+1}}\mathbb{I}[\log Y_i\ge \boldsymbol{X}_i^\top\check{\boldsymbol{\beta}} (\tau_r) ]dH(u) \right) = 0,
\end{equation}
 where $N_i(t)=\mathbb{I}( \log Y_i \le t, \delta_i=1)$ and $H(u) = -\log (1-u)$;  see \cite{peng2008survival}.

\begin{figure}[!ht]
    \centering
    \includegraphics[width = 0.75\textwidth]{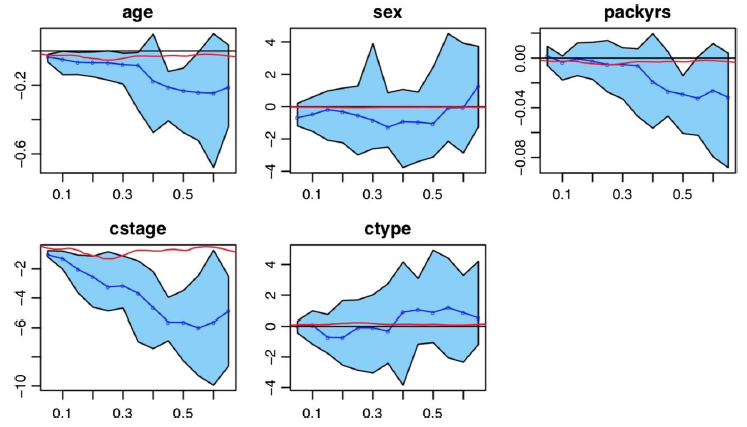}
    \caption{Example censored quantile regression analysis on $n = 153$ patients from the BLCSC study. 
    Estimated local quantile measures \citep{portnoy2003censored} from a Cox model are shown in red, and the reference $\beta = 0$ is given in black.
    The figure is a permitted use of Figure 2 in \cite{hong2019quantile}, published in {\it Precision Clinical Medicine}.}
    \label{fig:3}
\end{figure}

As a concrete example, with a subset of 153 patients from the BLCSC study, \cite{hong2019quantile} fit a censored quantile regression model that linked the conditional quantile of overall survival to age (years), sex (0:~female; 1:~male), pack-years, cancer type (0:~adenocarcinoma; 1:~non-adenocarcinoma), and cancer stage (0:~stage one; 1:~stage two or above). {\bf Figure \ref{fig:3}} displays the point estimates (blue curves) of the quantile-specific regression coefficients and their 95\% confidence intervals (lighter blue shaded regions).
 
While methods have been proposed to deal with variable selection for high-dimensional censored quantile regression (HDCQR), including penalized quantile regression \citep{wang2013variable}, adaptive penalized quantile regression \citep{zheng2013adaptive}, model-free variable screening \citep{he2013quantile}, and stochastic integral-based estimating equations \citep{zheng2018high}, none could draw statistical inference with HDCQR. \cite{belloni2019valid} provided post-selection inference in high-dimensional quantile regression at some fixed quantile levels; however, the method cannot handle censored outcomes. 
 
To address this issue, \cite{fei2021inference} proposed a Fused-HDCQR method, which utilizes a variable selection procedure for HDCQR \citep{zheng2018high} to reduce the dimension of the data, and applies partial regression to estimate the effect of each predictor, regardless whether it is selected. Estimates are aggregated based on multiple data splits and selections. Specifically, when $p>n$,  Fused-HDCQR adapts the SPARES procedure and fits low dimensional CQR's using Equation (\ref{cqrlow}).  With $B$ random sample splittings, these estimates, denoted by $\tilde{\beta}_j^b, b=1, \ldots, B$, are aggregated to form the Fused-HDCQR estimator:
\[
\begin{gathered}
	\hat{\beta}_j(\tau_k) = \frac{1}{B} \sum_{b=1}^{B} \tilde{\beta}_j^b(\tau_k),\;\tau_k \in \Gamma_m  \\ \hat{\beta}_j(\tau) = \hat{\beta}_j(\tau_k),\; \tau_{k-1}\le \tau < \tau_k,\, k=1,\ldots,m,
\end{gathered}
\]
and a {functional delta method \mbox{\citep{van2000asymptotic}}} can be applied to  estimate the variance of  $\hat{\beta}_j(\tau_k)$. The Fused-HDCQR procedure involves repeated fitting of low dimensional regressions, which is computationally feasible and can estimate and conduct hypothesis tests for the heterogeneous effects of various risk factors.

The use of Fused-HDCQR is illustrated with the BLCSC data by studying the differential impacts of  genetic variants on different quantiles of survival times. For example, \cite{fei2021inference}  focused on 2,002 candidate SNPs residing in 14 well-known lung cancer related genes and investigated how each SNP played a different role among high- (i.e., lower quantiles of overall survival) versus low-risk (i.e., higher quantiles of overall survival) cancer survivors. With the Fused-HDCQR approach, the estimated coefficient of active smoking ranged from $-0.42$ ($p=0.0011$)  to $-0.53$ ($p=0.0005$) as $\tau$ changed from $0.2$ to $0.5$, and then increased to $-0.31$ ($p=0.038$) as $\tau$ changed to $0.7$, suggesting that active smoking might be more harmful to the high- or median-risk groups than the low-risk group of patients. The results resonated with the strong need to develop effective smoking cession programs among high-risk populations \citep{barbeau2006results}. Further, SNP AX.37793583\_T remained significant throughout $\tau=0.2$ to $\tau=0.7$, however, its estimated coefficient decreased from $2.75$ ($\tau=0.2$) to $1.39$ ($\tau=0.7$), indicating its heterogeneous impacts on survival, i.e., stronger protective effect at lower quantiles and vice versa, which could not be detected using traditional Cox models \citep{fei2021inference}.


\section{MACHINE LEARNING TECHNIQUES FOR SURVIVAL PREDICTION}

Significant work has gone into the development of machine learning algorithms that can accommodate survival data.~These nonparametric learning approaches can handle non-linear relationships or higher-order interaction that would otherwise be costly in classical methods, and can improve accuracy in prediction for survival outcomes.

\subsection{Support Vector Machines}

Support vector machines (SVMs) fall under the {\it supervised learning} family \citep{vapnik1995support, noble2006support} and seek to find a hyperplane that provides maximal separation between groups ({\bf Figure \ref{fig:4}}). Specifically, consider a binary outcome $Y_i \in \{-1, 1\}$ for each individual $i$ with a corresponding $p$-dimensional covariate vector, $\boldsymbol{X}_i$. The goal of SVM is to identify a hyperplane, $H(\psi, a) = \{ \bv \in \mathbb{R}^p | \langle \psi, \bv \rangle + a = 0\}$, separating these two groups so that the margin, $2/|| \psi||$, can be maximized, where $\psi \in \mathbb{R}^p$ is the slope vector, and $\langle\cdot,\cdot\rangle$ denotes the inner product ({\bf Figure \ref{fig:4}}). Often, the two classes may not be separable in the original feature space within $\mathbb{R}^p$, and we use $\bF(\cdot)$ to map the original predictors to a higher dimensional space where  the outcomes can be distinguished,  in which case,  the hyperplane to deal with is $H(\psi, a) = \{ \bv \in \mathbb{R}^p | \langle \psi, \bF(\bv) \rangle + a = 0\}$ and, with slight overuse of notation, the dimension of $\psi$ is the same as that of $\bF(\bv)$. In practice, $\bF(\cdot)$ does not have to be obtained explicitly and $\langle \psi, \bF(\bv) \rangle$ can be calculated by using a reproducing kernel  \citep{wahba1999support}. We further introduce a slack variable, $\xi_i=[1- Y_i \{\langle \psi, \bF(\boldsymbol{X}_i)\rangle + a\}]_{+}$, 
to dictate the degree to which the $i$th data point is misclassified, as illustrated in {\bf Figure \ref{fig:4}}.

\begin{figure}[!ht]
    \centering
      \tikzset{
    leftNode/.style={circle, minimum width = .5ex, fill = white, draw},
    rightNode/.style={minimum width = .5ex, fill = cyan!20, thick, draw},
    rightNodeInLine/.style={solid,minimum width=.7ex, fill=cyan!20,thick,draw=black},
    leftNodeInLine/.style={solid,circle,minimum width=.7ex, fill=white,thick,draw},
    leftNodeMiss/.style={solid,circle,minimum width=.7ex, fill=headcolor!70,thick,draw=black}
  }
  \begin{tikzpicture}[
        scale=1.5,
        important line/.style={thick}, dashed line/.style={dashed, thin},
        every node/.style={color=black},
    ]
    
    \draw[help lines, color=gray!25, dashed] (0,0) grid (4,4);
    \draw[->,ultra thick] (0,0)--(4,0) node[right]{$v_1$};
    \draw[->,ultra thick] (0,0)--(0,4) node[above]{$v_2$};
    
    \draw[dashed line, thick]
       (0,1) coordinate (sls) -- (3,4) coordinate (sle)
       node [above right, anchor=east] {$\langle \psi, \bv \rangle + a = 1\ $}
       node[solid,circle,minimum width=2.8ex,fill=none,thick,draw] (name) at (2,3){}
       node[leftNodeInLine] (name) at (2,3){}
       node[solid,circle,minimum width=2.8ex,fill=none,thick,draw] (name) at (1,2){}
       node[leftNodeInLine] (name) at (1,2){};

    \draw[important line, thick]
       (0,0) coordinate (lines) -- (4,4) coordinate (linee)
       node [above right] {$\langle \psi, \bv \rangle + a = 0$};

    \draw[dashed line, thick]
       (1,0) coordinate (ils) -- (4,3) coordinate (ile)
       node [above right] {$\langle \psi, \bv \rangle + a = -1$}
       node[leftNodeMiss] (M1) at (3.5,2){}
       node[leftNodeMiss] (M2) at (1.5,1.8){}
       node[solid,minimum width=2.4ex, minimum height = 2.4ex, fill=none,thick,draw] (name) at (2.5,1.5){}
       node[rightNodeInLine] (name) at (2.5,1.5){};
      
    \draw node[] (S1) at (1.0,3.5){$\xi = 0$};
    \draw node[] (S1) at (1.7,2.1){$\xi < 1$};
    \draw node[] (S1) at (3.9,2.2){$\xi > 1$};
    \draw node[] (S1) at (3.5,1.0){$\xi = 0$};
      
    \draw node[leftNode]  (L1) at (4.5,2.0){};
    \draw node[anchor = west]() at (4.7,2.0){Class = 1};
    \draw node[rightNode] (L2) at (4.5,1.6){};
    \draw node[anchor = west]() at (4.7,1.6){Class = -1};
    
    \draw node[leftNodeInLine]  (L3) at (4.5,1.2){};
    \draw node[solid, circle, minimum width=2.8ex, minimum height = 2.8ex, fill=none,thick,draw] () at (4.5,1.2){};
    \draw node[anchor = west] () at (4.7,1.2){Support Vector};
    \draw node[rightNodeInLine] (L4) at (4.5,0.8){};
    \draw node[solid,minimum width=2.4ex, minimum height = 2.4ex, fill=none,thick,draw] () at (4.5,0.8){};
    \draw node[anchor = west]() at (4.7,0.8){Support Vector};
    
    \draw node[leftNodeMiss]  (L5) at (4.5,0.4){};
    \draw node[anchor = west]() at (4.7,0.4){Misclassified Point};
    
    \foreach \Point in {(.9,2.4), (1.3,2.5), (1.5,3.1), (2,3), (1,2.9)}{
      \draw \Point node[leftNode]{};
    }

    \foreach \Point in {(2.9,1.4), (2.3,.5), (3.3,.1), (2,0.9), (2.5,1)}{
      \draw \Point node[rightNode]{};
    }
    
  \end{tikzpicture}
    \caption{A support vector machine to distinguish binary outcomes with two-dimensional covariates  $\bv=(v_1, v_2)^\top$ by a linear separating line. The solid line represents the optimal hyperplane separating the data, while the dotted lines denote the maximal margin defined by the support vectors (encircled nodes) for one group (white circles) versus the other (cyan squares). Misclassified points are labeled in red, with corresponding magnitudes for slack variables, $\xi$.} 
    \label{fig:4}
\end{figure}

SVMs have been extended to model continuous time-to-event data, which are prone to censoring, by predicting the survival time to be $\langle \psi, \bF(\boldsymbol{X}_i)\rangle + a$.~\cite{van2007support} formulated the {\it survival SVM} based on the rank concordance between the prediction and observed survival time, $Y_i$, among  comparable individuals in the presence of censoring. Specifically, they introduced a {\it comparability indicator}, $v_{ij}= \delta_i \mathbb{I}( Y_i < Y_j)$, such that the ordering of the observed survival times for subjects $i,j$ can only be determined when $v_{ij}=1$.~For a comparable pair with $v_{ij}=1$, a concordance in rank is reached if and only if $ \langle \psi, \bF(\boldsymbol{X}_j)\rangle  - \langle \psi, \bF(\boldsymbol{X}_i)\rangle  >0$. Allowing varying degrees of pairwise slacks,
i.e., when $ \langle \psi, \bF(\boldsymbol{X}_j)\rangle  - \langle \psi, \bF(\boldsymbol{X}_i)\rangle  \le 0$ with $v_{ij}=1$,  across comparable  pairs,  \cite{van2007support} proposed to solve
\begin{align}
    \begin{split} 
        \min _{\psi, \xi} & \frac{1}{2}\|\psi\|^{2}+\gamma \sum_{(i,j):Y_i< Y_j} v_{i j} \xi_{i j} \nonumber \\
        \text { subject to } & \left\langle\psi, \bF\left(\boldsymbol{X}_{j}\right)\right\rangle-\left\langle\psi, \bF\left(\boldsymbol{X}_{i}\right)\right\rangle \geq -\xi_{i j},  \nonumber \\
        \text { and } & \xi_{i j} \geq 0, i, j=1, \ldots, n,
    \end{split}
\end{align}
where $\xi_{ij}$'s are pair-specific slacks, whose summation is to be minimized, and $\gamma > 0$ is a regularization parameter controlling the maximal margin and misclassification penalties. This formulation can be shown to maximize the Harrell rank-based concordance index (C-index) \citep{harrell1982evaluating}. Hence, it is termed the {\it rank-based} SVM approach for survival data and does not estimate the ``intercept" $a$. An alternative {\it regression} approach \citep{shivaswamy2007support} aimed to find a prediction, $\langle \psi, \bF(\boldsymbol{X}_i)\rangle + a$, for continuous survival times, by identifying a hyperplane that best fit the data that are subject to censoring \citep{smola2004tutorial}:  
\begin{align} \label{eq:svmreg}
    \begin{split}
        \min_{\psi, a, \xi_{,}, \xi^{*}} & \frac{1}{2}\|\psi\|^{2}+\gamma \sum_{i=1}^{n}\left(\xi_{i}+\xi_{i}^{*}\right)  \nonumber \\
        \text { subject to } & Y_{i}-\left\langle\psi, \bF\left(\boldsymbol{X}_{i}\right)\right\rangle-a \leq \xi_{i}, \\
        & \delta_{i}\left(\left\langle\psi, \bF\left(\boldsymbol{X}_{i}\right)\right\rangle+a-Y_{i}\right) \leq \xi_{i}^{*}, \nonumber \\
        \text { and } & \xi_{i}, \xi_{i}^{*} \geq 0.
    \end{split}
\end{align}

With censoring indicators incorporated into the constraints, the formulation utilizes available information from both censored and non-censored observations. To make full use of the strengths of both approaches, \cite{van2011support} and \cite{polsterl2015fast} further proposed {\it hybrid} approaches, combining the penalties imposed by both methods.

\subsection{Tree-Based Methods}
While SVMs are adept at estimating non-linear relationships, they do not scale well for large datasets and often under-perform when the outcomes are noisy. Also there may be no clear interpretations for classifying data points above or below the estimated hyperplane \citep{7860040}. Decision trees are an alternative for classifying patients that provide an intuitive interpretation of the hierarchical relationships between predictors. Broadly, classification and regression trees (CART) is an umbrella term for a set of {\it recursive partitioning} algorithms, which predict the group membership (classification) or target value (regression) for an observation based on a set of binary decision rules ({\bf Figure \ref{fig:5}}). 

\cite{gordon1985tree} first presented survival trees, and  \cite{ciampi1986stratification, ciampi1987recursive} solidified the notion and established splitting criteria based on the log-rank and likelihood ratio test statistics, respectively, gaining predictive accuracy and interpretability. A recursive partitioning algorithm for generating a survival tree is given as follows.
\begin{enumerate}
    \item Discretize each covariate to be a binary variable (categorical variables with $m$ levels are expressed as $m-1$ dummy variables).
    
    \item For every binary covariate, $X_j, \, j=1, \ldots, p$, compute the log-rank statistic to test the difference between the survival curves for the two groups defined by $X_j$.

    \item Choose the covariate, say, $X_{j^*}$, with the largest significant test statistic and partition the full sample (i.e., the {\it root node}) into two groups ({\it child nodes}) based on $X_{j^*}$.  

    \item Repeat steps 2-3 for each subset ({\it child node}) until reaching the {\it terminal nodes}, that is, no covariates produce a significant test statistic and there are enough events (exceeding a prespecified number) in each {\it terminal node}.
\end{enumerate}
The resulting  terminal nodes split the original sample into distinct groups, who are deemed more homogeneous within each group, and will output survival estimates via Kaplan-Meier estimation in each group.
Further variations in splitting are based on metrics that accommodate censored data and by either minimizing within-node homogeneity or maximizing between-node heterogeneity. For example, these metrics  can be Martingale residuals \citep{therneau1990martingale} or deviance residuals \citep{leblanc1992relative}.  With an established splitting criterion, to select a final tree, either a full survival tree is `grown' and `pruned' or a stopping rule is applied in backward or forward selection \citep{bou2011review}.

\tikzset{
  red arrow/.style={
    midway,headcolor,sloped,fill, minimum height=3cm, single arrow, single arrow head extend=.5cm, single arrow head indent=.25cm,xscale=0.3,yscale=0.15,
    allow upside down
  },
  black arrow/.style 2 args={-stealth, shorten >=#1, shorten <=#2},
  black arrow/.default={1mm}{1mm},
  tree box/.style={draw, rounded corners, inner sep=1em},
  node box/.style={white, draw=black, text=black, rectangle, rounded corners},
}

\begin{figure}[!ht]
    \centering
    \tikzstyle{block} = [circle, draw=headcolor!70, fill=headcolor!70, 
    text width=1em, text centered, rounded corners, minimum height=1em]
    \tikzstyle{line} = [draw, -]
    \tikzstyle{cloud} = [draw, text centered, circle,fill=cyan!20, node distance=0cm, text width=1em,
    minimum height=0.0em]
    \begin{tikzpicture}[node distance =0cm, auto]
    \small
    \centering
    \node [block] (init) {\scriptsize{$X_1$}};
    \node [cloud, left of=init, below of=init, node distance=1.5cm] (end1) {};
    \node [below of=end1, node distance=0.5cm] (lab1) {\scriptsize{Terminal Node 1}};
    \node [block, right of=init, below of=init, node distance=1.5cm] (dec1) {\scriptsize{$X_2$}};
    \node [cloud, right of=dec1, below of=dec1, node distance=1.5cm] (end2) {};
    \node [below of=end2, node distance=0.5cm] (lab2) {\scriptsize{Terminal Node 2}};
    \node [block, left of=dec1, below of=dec1, node distance=1.5cm] (dec2) {\scriptsize{$X_3$}};
    \node [cloud, left of=dec1, below of=dec2, node distance=1.5cm] (end3) {};
    \node [below of=end3, node distance=0.5cm] (lab3) {\scriptsize{Terminal Node 3}};
    \node [block, right of=dec1, below of=dec2, node distance=1.5cm] (end4) {};
    \node [below of=end4, node distance=0.5cm] (lab4) {\scriptsize{Terminal Node 4}};
    \path [line] (init) -- (end1)
        node [midway, above, rotate = 45] {\scriptsize{$X_1 = 1$}};
    \path [line] (init) -- (dec1) 
        node [near start, rotate = -45] {\scriptsize{$X_1 = 0$}} ;
    \path [line] (dec1) -- (dec2) 
        node [midway, above, rotate = 45] {\scriptsize{$X_2 = 1$}};
    \path [line] (dec1) -- (end2) 
        node [near start, rotate = -45] {\scriptsize{$X_2 = 0$}};
    \path [line] (dec2) -- (end3)
        node [midway, above, rotate = 45] {\scriptsize{$X_3 = 1$}};
    \path [line] (dec2) -- (end4)
        node [near start, rotate = -45] {\scriptsize{$X_3 = 0$}};
    \path[-latex, headcolor!70] ([yshift=-0.5ex, xshift= 5]init.south) edge ([yshift=1ex, xshift=-3]dec1.west);
    \path[-latex, headcolor!70] ([yshift=-0.5ex, xshift=-5]dec1.south) edge ([yshift=1ex, xshift= 3]dec2.east);
    \path[-latex, headcolor!70] ([yshift=-0.5ex, xshift= 5]dec2.south) edge ([yshift=1ex, xshift=-3]end4.west);
    \end{tikzpicture}
    \caption{Illustration of a classification tree with three binary covariates that yields four terminal nodes. 
    }
    \label{fig:5}
\end{figure}

\subsection{Ensemble Learners}

While survival trees provide a fast and intuitive means of studying hierarchical relationships of predictors with outcomes, they are prone to over-fitting and high variability  \citep{hu2018personalized, steingrimsson2016doubly}. Ensemble learners  overcome the instability issues by using  techniques such as  {\it bagging}, {\it boosting}, and {\it random forests}.

\subsubsection{Bagging} 

Bootstrap aggregation or {\it bagging} refers to a means of training an ensemble learner by resampling the data with replacement, training weak learners (e.g., individual survival trees) in parallel, and combining these  results over the multiple {\it bootstrapped} samples \citep{breiman1996bagging}. It has three steps.
\begin{enumerate}
    \item {\bf Bootstrapping}: Resample from the original data of size $n$ {\it with replacement} to form a new sample also of size $n$, and  obtain `$B$' such samples. 
    
    \item {\bf Parallel Training}: With each bootstrap sample, $b = 1, \ldots, B$, independently train the weak learners in parallel.
    
    \item {\bf Aggregation}: Combine the $B$ individual predictions by averaging over them or by taking a majority vote.
\end{enumerate}

Bagging for survival trees was first proposed by \cite{hothorn2004bagging}; in contrast to bagging for classification trees, aggregation is done by averaging  survival predictions, rather than a `majority vote.' Each survival tree is grown so that every terminal node has enough events, which are used to predict the survival function node-wise at each terminal node. Then, for any newcomer, the predictions are averaged over the individual trees to yield an ensemble prediction of their survival function  ({\bf Figure \ref{fig:6}}). 



\begin{figure}[!ht]
    \centering
    \scalebox{0.4}{
    \begin{forest}
        for tree={l sep=3em, s sep=3em, anchor=center, inner sep=0.7em, fill=cyan!20, draw=black, circle, where level=2{no edge}{}}
        [
        Training Sample, node box
        [Bootstrap Resampling, node box, alias=bagging, above=4em
        [,headcolor!70,alias=a1[[,alias=a2][]][,headcolor!70,edge label={node[above=1ex,red arrow]{}}[[][]][,headcolor!70,edge label={node[above=1ex,red arrow]{}}[,headcolor!70,edge label={node[below=1ex,red arrow]{}}][,alias=a3]]]]
        [,headcolor!70,alias=b1[,headcolor!70,edge label={node[below=1ex,red arrow]{}}[[,alias=b2][]][,headcolor!70,edge label={node[above=1ex,red arrow]{}}]][[][[][,alias=b3]]]]
        [~~$\dots$~,scale=2,no edge,fill=none,draw = none, yshift=-4em]
        [,headcolor!70,alias=c1[[,alias=c2][]][,headcolor!70,edge label={node[above=1ex,red arrow]{}}[,headcolor!70,edge label={node[above=1ex,red arrow]{}}[,alias=c3][,headcolor!70,edge label={node[above=1ex,red arrow]{}}]][,alias=c4]]]]
        ]
        \node[tree box, fit=(a1)(a2)(a3)](t1){};
        \node[tree box, fit=(b1)(b2)(b3)](t2){};
        \node[tree box, fit=(c1)(c2)(c3)(c4)](tn){};
        \node[below right=0.5em, inner sep=0pt] at (t1.north west) {Tree 1};
        \node[below right=0.5em, inner sep=0pt] at (t2.north west) {Tree 2};
        \node[below right=0.5em, inner sep=0pt] at (tn.north west) {Tree B};
        \path (t1.south west)--(tn.south east) node[midway,below=4em, node box] (mean) {Aggregation};
        \node[below=3em of mean, node box] (pred) {Prediction};
        \draw[black arrow={5mm}{4mm}] (bagging) -- (t1.north);
        \draw[black arrow] (bagging) -- (t2.north);
        \draw[black arrow={5mm}{4mm}] (bagging) -- (tn.north);
        \draw[black arrow={5mm}{5mm}] (t1.south) -- (mean);
        \draw[black arrow] (t2.south) -- (mean);
        \draw[black arrow={5mm}{5mm}] (tn.south) -- (mean);
        \draw[black arrow] (mean) -- (pred);
    \end{forest}
    }
    \caption{Ensemble learning with bootstrap aggregation (bagging) for survival trees.}
    \label{fig:6}
\end{figure}

\subsubsection{Boosting} In a similar vein, {\it boosting}  trains a series of weak learners with the goal of aggregating them into a better ensemble learner \citep{buhlmann2007boosting}. \cite{hothorn2006survival} proposed a  gradient boosting algorithm for survival settings. Consider a mortality risk prediction  based on covariates, $\boldsymbol{X}_i$. For an $M$-step gradient boosting algorithm, a prediction, ${\cal F}_m(\boldsymbol{X}_i)$, is made at each step, say $m=1,\ldots,M$, based on a previous prediction, ${\cal F}_{m-1}(\boldsymbol{X}_i)$, and an additional weak learner $f_{m}(\boldsymbol{X}_i),$ which is the projection of the ``residual error" of ${\cal F}_{m-1}(\boldsymbol{X}_i)$ to the space spanned by $\boldsymbol{X}_i$,
$${\cal F}_m(\boldsymbol{X}_i) = {\cal F}_{m-1}(\boldsymbol{X}_i) + w_m  f_m(\boldsymbol{X}_i),$$
where $ 0<w_m \le 1$ (e.g., $w_m=0.1$) is the step size,  
the residual error refers to the gradient of the loss function, e.g., the negative log partial likelihood function in a survival setting,  evaluated at
${\cal F}_{m-1}(\boldsymbol{X}_i)$, and the number of steps, $M$, can be viewed as a tuning parameter.

Boosting has two notable differences from bagging. First, boosting trains  weak learners sequentially, updating the weights placed on learners iteratively, whereas
in bagging individual weak learners such as survival trees are trained independently and in parallel, which are aggregated via majority voting or averaging. Second,  boosting is applicable to settings where learners have low variability and high bias, as the performance is improved by  redistributing the weights. In contrast,
bagging is often applied when  individual learners exhibit high variability, but low bias, as it reduces variations arising from individual trees.

\subsubsection{Random Forests} 

Yet another class of ensemble learners are random forests \citep{breiman2001random}, which, like bagging, aggregate predictions from individual trees  generated over bootstrap resampled datasets. However, differing from bagging,  random forests randomly select a subset of features, say $p'<p$ features, when generating each tree and use them for the individual tree's growth. By doing so, random forests reduce correlations among individual trees, leading to gains in accuracy \citep{breiman2001random}. The choice of $p'$ is problem-specific, which can also be viewed as a tuning parameter. In survival settings, \cite{ishwaran2008random} aggregated the survival predictions arising from each tree by averaging the predicted cumulative hazard functions into an ensemble prediction. 

Further notable developments include \cite{ishwaran2011random}, which extended random survival forests to high dimensions by incorporating regularization, \cite{ishwaran2019standard}, which provided standard errors and confidence intervals for variable importance, and \cite{steingrimsson2019censoring}, which proposed censoring unbiased regression trees and ensembles.

\subsection{Deep Learning and Artificial Neural Networks}

Deep learning has emerged as a powerful tool for risk prediction.~This work stems from artificial neural networks that tried to mirror how the human brain functions \citep{rosenblatt1958perceptron}, wherein nodes (or {\it neurons}) are connected in a network as a weighted sum of inputs through a series of affine transformations and non-linear activations. 

A fully-connected, feed-forward artificial neural network is made up of $L$ layers, with $k_{l}$ neurons in the $l$th layer $(l=1, \ldots, L)$ ({\bf Figure \ref{fig:7}}). With an input, network predictions are made based on an $L$-fold composite function, $ f_{L} \circ f_{L-1} \circ \cdots \circ f_{1}(\cdot)$ with $(g\circ f)(\cdot) = g(f(\cdot))$.  At the $l$h layer, $f_l(\cdot)$, is defined as
$$f_{l}(\bv)=\sigma_{l}\left(\mathbf{W}_{l} \bv+\bb_{l}\right) \in \mathbb{R}^{k_{l}},$$ 
where  $\bv$ is  a $k_{l-1} \times 1$ input vector fed from the $(l-1)$th layer,
$\sigma_{l} (\cdot): \mathbb{R}^{k_{l}} \rightarrow \mathbb{R}^{k_{l}}$ is an activation function, $\mathbf{W}_{l}$ is a $k_{l} \times k_{l-1}$ weight matrix,  $\bb_{l}$ is a $k_{l} \times 1$ bias vector, and the $0$th layer is the input layer. Typical choices of  $\sigma_{l} (\cdot)$ include the sigmoid function or the rectified linear unit activation function (ReLU), that is, $ \sigma_{l}(\bb)=\max (0,\bb)$, where
$\bb \in \mathbb{R}^{k_{l}}$ and $\max(0, \cdot)$  operates component-wise.

\begin{figure}[!ht]
    \centering
    \tikzset{
        every neuron/.style={circle, draw, fill=cyan!20, minimum size=0.5cm},
        neuron missing/.style={draw=none, fill=none, scale=1,text height=0.333cm, execute at begin node=\color{black}$\vdots$},
    }
    \begin{tikzpicture}[x=1.5cm, y=1.5cm, >=stealth, scale = 0.5]
        \foreach \m/\l [count=\y] in {1,2,3,missing,4}
            \node [every neuron/.try, neuron \m/.try] (input-\m) at (0,2.5-\y) {};
        \foreach \m [count=\y] in {1,missing,2}
            \node [every neuron/.try, neuron \m/.try] (hidden-\m) at (2,2-\y*1.25) {};
        \foreach \m [count=\y] in {1,missing,2}
            \node [every neuron/.try, neuron \m/.try, fill=none ] (output-\m) at (4,1.5-\y) {};
        \foreach \l [count=\i] in {1,2,3,{{k_0}}}
            \draw [<-] (input-\i) -- ++(-1,0)
                node [above, midway] {$I_\l$};
        \foreach \l [count=\i] in {1,{{k_1}}}
            \node [above] at (hidden-\i.north) {$H_\l$};
        \foreach \l [count=\i] in {1,{{k_2}}}
            \draw [->] (output-\i) -- ++(1,0)
                node [above, midway] {$O_\l$};
        \foreach \i in {1,...,4}
            \foreach \j in {1,...,2}
                \draw [->] (input-\i) -- (hidden-\j);
        \foreach \i in {1,...,2}
            \foreach \j in {1,...,2}
                \draw [->] (hidden-\i) -- (output-\j);
        \foreach \l [count=\x from 0] in {input, hidden, output}
            \node [align=center, above] at (\x*2,2) {\l \\ layer};
    \end{tikzpicture}
    \caption{Diagram of a feed-forward, fully-connected two-layer artificial neural network, including the hidden (1st)  and output (2nd) layer. The input ($0$-th) layer is not counted as a real neural network layer.}
    \label{fig:7}
\end{figure}

 For survival prediction, several deep learning approaches have emerged, beginning with the seminal work of \cite{faraggi1995neural}, which adopted a fully-connected, feed-forward neural network to extend the Cox model to nonlinear predictions.
 Other feed-forward neural networks \citep{liestbl1994survival, brown1997use,  biganzoli1998feed,eleuteri2003novel} used the survival status as a training label, and output predicted survival probabilities. Further developments have been made in Bayesian networks \citep{
 bellazzi2008predictive, lisboa2003bayesian, fard2016bayesian}, convolutional neural networks \citep{yao2017deep, katzman2016deep, katzman2018deepsurv, ranganath2016deep}, and recurrent neural networks \citep{
 yang2018spatio}.


\section{PREDICTION FOR COMPETING AND SEMI-COMPETING RISKS}

Many survival processes in real applications involve multiple competing events. Risk prediction in these settings is an up-and-coming field with many potential developments. We focus on two common competing event settings, i.e.,  competing and semi-competing risks.

\subsection{Competing Risks}

{In a competing risk setting,  observing an event type, labeled by $c \in \{1, \ldots, K\}$, effectively eliminates the chance of observing 
other event types happening to the same individual \mbox{\citep{young2020causal}}.} For example, when studying the survival of  patients with cancer, competing events can be cancer-related death ($c=1$) or death by cardiac disease ($c=2$) ({\bf Figure \ref{fig:8}});  an individual cannot die of  cardiac disease once they have died of cancer, and vice versa. For characterizing the risk of competing events, there are two commonly used statistical metrics, namely, the cause-specific hazard and the subdistribution hazard, which target different counterfactual scenarios. The former describes the risk under hypothetical elimination of competing events, while the latter is about the observable risk without elimination of any competing events
\citep{rudolph2020causal}. 


\begin{figure}[!ht]
    \centering
    \begin{tikzpicture}[x=1cm,y=0.4cm]
        \draw[-latex] (0,0)--(9,0) node [right] {Time}; 
        \draw[-] (1,0)--(1,6) node [above] {}; 
        \draw[-] (8,0)--(8,6) node [above] {}; 
        \draw[|-] (1.5,5)--(4,5) node [right] {\hskip 2ex Cancer Death $(c=1)$};
        \draw[|-] (2,3)--(3,3) node [right] {\hskip 2ex Cardiac Death $(c=2)$};
        \draw[|-] (4,1)--(8,1) node [right] {\hskip 2ex Censoring};
        \node[] at (1, -1) {Study Start};
        \node[] at (8, -1) {Study End};
        \node[] at (0,5) {Patient 1};
        \node[headcolor] at (4,5) {\huge $\times$};
        \node[] at (0,3) {Patient 2};
        \node[cyan] at (3,3) {\huge $\times$};
        \node[] at (0,1) {Patient 3};
        \node[white] at (8,1) {\huge \textbullet};
        \node[] at (8,1) {\huge $\circ$};
    \end{tikzpicture}
    \vspace{2ex}
    \caption{Schematic of observation times for three example patients with competing risks;  {\color{headcolor}{$\times$}}: cancer death;  
    {\color{cyan}{$\times$}}: cardiac death;
    {$\circ$}: censoring.
    }
    \label{fig:8}
\end{figure}

Several authors \citep{lau2009competing, koller2012competing} have stated that the subdistribution hazard is useful for predicting  the probability of having an event of a type of interest by a given time, termed the  cumulative incidence function (CIF),   which reflects an individual’s actual risks and prognosis. In the following, we  focus on the subdistribution  hazard, which is derived from CIF, i.e., $F_c(t)=\operatorname{Pr}(T_i < t,  {\cal C}_i=c)$, where ${\cal C}_i$ marks the event type for subject $i$. Specifically, for  each event type $c=1, \ldots, K$, it is
defined as 
\[
  \lambda_{c}(t)=\lim _{\Delta  \rightarrow 0} \frac{\operatorname{Pr}\left(t \le T_i<t+\Delta,  {\cal C}_i=c \mid T_i \ge t  \cup \{ T_i < t  \wedge {\cal C}_i \neq c \} \right)} {\Delta}
  = \frac{d F_c(t)/dt}{1- F_c(t)},
\]
which denotes the instantaneous risk of failure from event type $c$ among those who have not experienced this type of event. That is, the risk set at
$t$ includes those who are event free as well as those who have experienced a competing event (other than type $c$)  by $t$. The subdistribution hazard model \citep{fine1999proportional} links a subdistribution hazard function to covariates via 
\begin{equation} \label{subhar}     
    \lambda_{c}(t|\boldsymbol{X}_{i})=\lambda_{0 c}(t) \exp (\boldsymbol{X}_{i}^\top\boldsymbol{\beta}),
\end{equation}
where $\lambda_{0 c}(t)$ is the baseline subdistribution hazard function for event type $c$, and  $\boldsymbol{\beta}$ specifies the effect of $\boldsymbol{X}_{i}$ on the probability of event $c$ occurring over time. In fact, model (\ref{subhar}) implies that $1-F_c(t|\boldsymbol{X}_{i})
= \{1-F_{0c}(t)\}^{ \exp(\boldsymbol{X}_{i}^\top\boldsymbol{\beta})}$, where
$F_c(t|\boldsymbol{X}_{i})$ and $F_{0c}(t)$ are the CIF given $\boldsymbol{X}_{i}$ and the baseline CIF, respectively.

  {With high-dimensional predictors,   several authors} 
  \citep{kawaguchi2019fast,ha2014variable,ahn2018group}  {proposed regularized subdistribution hazard models for variable selection, and \mbox{\cite{hou2019inference}} further performed inference using a one-step debiased LASSO estimator.} For prediction, several deep learning works for competing risks have been proposed based on CIFs. For example, DeepHit \citep{lee2018deephit} developed a multi-task network to nonparametrically estimate $F_c(t|\boldsymbol{X}_{i})$ for $c=1, \ldots, K$.  The network is trained to minimize a loss function, which is constructed based on the joint distribution of the first hitting time for competing events of each subject, while  ensuring the concordance of estimates across subjects \citep{harrell1982evaluating}, that is, a patient who died at a given time should have a higher risk at that time than a patient who survived longer.
 Dynamic DeepHit \citep{lee2019dynamic} further incorporated longitudinal information for dynamic predictions. Other approaches have included DeepCompete \citep{aastha2020deepcompete}, as well as a hierarchical approach to multi-event survival analysis \citep{tjandra2021hierarchical}.

\subsection{Semi-Competing Risks}


Semi-competing risk problems, a variant of competing risk problems, have commonly been encountered in clinical studies. By {\it semi}-competing, we mean that the occurrence of one event, i.e., a {\it non-terminal} event, is subject to the occurrence of another {\it terminal} event, but not vice versa ({\bf Figure \ref{fig:9}}). As the non-terminal event (e.g., cancer progression) is often a strong precursor to the terminal event (death), semi-competing events are often related and, hence, the terminal event may informatively censor the non-terminal event \citep{jazic2016beyond}. To overcome such informative censoring, researchers either consider only the terminal event (i.e., mortality) or a composite outcome such as {\it progression-free survival}, that is, time to progression or death, whichever comes first. 

What is lacking here is how to model a predictor's potentially different roles in disease progress and death, while utilizing the crucial information about the {\it sojourn} time between progression and death. Even in settings where the non-terminal and terminal event times are only modestly correlated, failing to acknowledge this sojourn time may lead to incorrect inference or biased predictions \citep{crilly2021predicting}. 

\begin{figure}[!ht]
    \centering
     \begin{tikzpicture}[x=1cm,y=0.4cm]
     
         \draw[-latex] (0,0)--(9,0) node [right] {\small Time}; 
         \draw[-] (1,0)--(1,8) node [above] {}; 
         \draw[-] (8,0)--(8,8) node [above] {}; 
         \draw[|-] (1.5,7)--(7,7); 
         \draw[dotted] (8,7) node [right] {\hskip 1ex \small Non-Terminal Followed by Terminal Event};
         \draw[|-] (3,5)--(8,5) node [right] {\hskip 1ex \small Non-Terminal Event Observed Only};
         \draw[|-] (2,3)--(3,3); 
         \draw[dotted] (8,3) node [right] {\hskip 1ex \small Terminal  Precludes
         Non-Terminal Event};
         \draw[|-] (4,1)--(8,1) node [right] {\hskip 1ex \small None Observed (Censoring)};
 
         \node[] at (1, -1) {\small Study Start};
         \node[] at (8, -1) {\small Study End};
         
         \node[] at (0,7) {\small Patient 1};
         \node[cyan!20] at (4,7) {\huge $\bLozenge$};
         \node[] at (4,7) {\huge $\Diamond$};
         \node[headcolor] at (7,7) {\huge $\times$};
         \node[] at (0,5) {\small Patient 2};
         \node[cyan!20] at (5,5) {\huge $\bLozenge$};
         \node[] at (5,5) {\huge $\Diamond$};
          \node[white] at (8,5) {\huge \textbullet};
         \node[] at (8,5) {\huge $\circ$};
         \node[] at (0,3) {\small Patient 3};
         \node[headcolor] at (3,3) {\huge $\times$};
         \node[] at (0,1) {\small Patient 4};
         \node[white] at (8,1) {\huge \textbullet};
         \node[] at (8,1) {\huge $\circ$};
 
    \end{tikzpicture}
    \vspace{2ex}
    \caption{Schematic of four example patients with semi-competing risks; {\color{cyan!20}{$\bLozenge$}}: non-terminal event;
    {\color{headcolor}{$\times$}}: terminal event;
    {$\circ$}: censoring.
    }
    \label{fig:9}
\end{figure}

\subsubsection{The Illness-Death Model}

Central to the formulation of the semi-competing problem is the {\it illness-death model}, a compartment-type model for the rates at which individuals transition from being event-free (e.g., from time of diagnosis) to progression or to death
or from progression to death
\citep{
andersen2012statistical}. Letting $T_{i1}$,   $T_{i2}$, and $C_i$ denote the times to the non-terminal and terminal events, and the censoring time, respectively, we observe $
(Y_{i1},\ \delta_{i1},\ Y_{i2},\ \delta_{i2},\ \boldsymbol{X}_i),\, i=1,\ldots,n$, where $Y_{i2} = \min(T_{i2}, C_i)$, $\delta_{i2} = \mathbb{I}(T_{i2} \leq C_i)$, $Y_{i1} = \min(T_{i1}, Y_{i2})$, $\delta_{i1} = \mathbb{I}(T_{i1} \leq Y_{i2})$, and $\boldsymbol{X}_i$ is a $p$-vector of covariates. The hazards  for each subject
at $t$ (since diagnosis) are defined and modeled as

\scriptsize
\begin{align}
    \lambda_{1}\left(t \mid \gamma_{i}, \boldsymbol{X}_{i}\right) &=
\lim _{\Delta  \rightarrow 0}   \frac{ \operatorname{Pr}(  t \le T_{i1} < t + \Delta  | T_{i1} \ge t, T_{i2} \ge t, \gamma_i,\boldsymbol{X}_i)}{\Delta}=  
    \gamma_{i} \lambda_{01}\left(t\right) \exp \left\{h_1(\boldsymbol{X}_i)\right\},  \label{eq:condhaz1}\\
    \lambda_{2}\left(t \mid \gamma_{i},\boldsymbol{X}_{i}\right) &=
    \lim _{\Delta  \rightarrow 0}   \frac{ \operatorname{Pr}(  t \le T_{i2} < t + \Delta  | T_{i1} \ge t, T_{i2} \ge t, \gamma_i,\boldsymbol{X}_i)}{\Delta}=
    \gamma_{i} \lambda_{02}\left(t\right) \exp \left\{h_2(\boldsymbol{X}_i)\right\},  \label{eq:condhaz2}\\
    \lambda_{3}\left(t \mid t_{1}, \gamma_{i}, \boldsymbol{X}_{i}\right) &=
    \lim _{\Delta  \rightarrow 0}   \frac{ \operatorname{Pr}(  t \le T_{i2} < t + \Delta  | T_{i1} = t_1, T_{i2} \ge t, \gamma_i,\boldsymbol{X}_i)}{\Delta}=
    \gamma_{i} \lambda_{03}\left(t - t_{1}\right) \exp \left\{h_3(\boldsymbol{X}_i)\right\}, \label{eq:condhaz3}
\end{align}
\normalsize
where  $0<t_{1}<t$, and (\ref{eq:condhaz1})-(\ref{eq:condhaz3}),  respectively,  correspond to the transition  from diagnosis to progression prior to death,  from diagnosis to death prior to progression,  and from progression (that happens at $t_1$) to death \citep{haneuse2016semi}.  Here, $\gamma_{i} \stackrel{i.i.d}{\sim} \Gamma(1/\theta, 1/\theta)$  (i.e., both shape and rate are $1/\theta$ so that the mean  and variance are respectively 1 and $\theta$),  $i=1, \ldots, n,$ is a patient-specific {\it frailty} that models the dependence among these three transition processes within subject $i$, that is, a larger value of $\theta$ reflects a stronger dependence. In addition, $\lambda_{0g}\left(\cdot\right), g = 1,2,3,$ are the baseline hazard functions for the three state transitions, respectively, and $h_g(\cdot), g = 1,2,3,$ are log-risk functions which relate a patient's covariates to each potential transition. The $\lambda_{0g}$ functions can be taken to be Weibull functions or piecewise constant with jumps at the distinct observed event times. Given (\ref{eq:condhaz1})-(\ref{eq:condhaz3}), and by integrating out the frailty term, \cite{reeder2022penalized} derived the marginal likelihood based on  $n$ independent subjects as
\begin{align}\label{eq:obj}
    \begin{split}
        \mathcal{L}= \prod_{i=1}^n & \{\lambda_{1i}(Y_{i1})\}^{\delta_{i1}} \{\lambda_{2i}(Y_{i1})\}^{\left(1-\delta_{i1}\right) \delta_{i2}} \{\lambda_{3i}(Y_{i2}-Y_{i1})\}^{\delta_{i1} \delta_{i2}}\left(1+\theta^{-1}\right)^{\delta_{i1} \delta_{i2}} \\
        & \times\left[1+\theta^{-1}\left\{\Lambda_{1i}\left(Y_{i1}\right)+\Lambda_{2i}\left(Y_{i1}\right)+\Lambda_{3i}\left(Y_{i2}-Y_{i1}\right)\right\}\right]^{-\theta-\delta_{i1}-\delta_{i2}}, 
    \end{split}
\end{align}
\noindent where $\lambda_{gi} (s) =  \lambda_{0g}\left(s \right) \exp \left\{h_g(\boldsymbol{X}_i)\right\}$and $\Lambda_{gi}(t)=\int_{0}^{t} \lambda_{gi}(s) ds$ for $g = 1,2,3$. 


\subsubsection{A New Deep Learning Approach for Semi-Competing Risks} 
\begin{figure}[!ht]
    \centering
    \includegraphics[scale = 0.75]{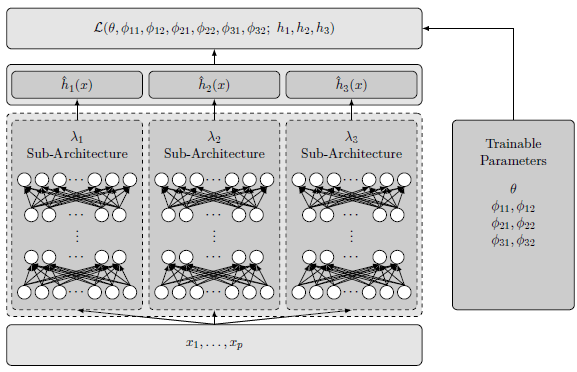}
    \caption{Architecture for the proposed semi-competing risk deep neural network.}
    \label{fig:10}
\end{figure}

We propose a multi-task deep neural network for semi-competing risks (DNN-SCR), by using Equation (\ref{eq:obj}) as the objective function with potentially high-dimensional covariates. DNN-SCR consists of three risk-specific sub-networks, respectively corresponding to the three possible state transitions, and a finite set of trainable parameters for specifying the baseline hazards (i.e., the $\phi$ parameters in {\bf Figure \ref{fig:10}}) if we specify Weibull baseline hazards, $\lambda_{0g}(s) = \phi_{g1} \phi_{g2} s^{\phi_{g2}-1}$ for $g=1, 2,3,$ in (\ref{eq:condhaz1})-(\ref{eq:obj}) as well as the dependence among the three transition processes (i.e., $\theta$ in {\bf Figure \ref{fig:10}}). As opposed to the classical models, we opt for flexible, nonparametric estimation of $h_g(\cdot), g=1,2,3,$ to better capture potential non-linear dependencies of covariates on semi-competing events and to maximize the predictive accuracy. 

In particular, we design three neural network sub-architectures to estimate the $h$ functions nonparametrically as outputs. For identifiability, we require
$h_g(\bold{0})=0, g=1, 2, 3$, where $\bold{0}$ is a $p \times 1$ vector of  0's. Each sub-network is a fully-connected feed-forward neural network with ReLU activation functions and a linear activation in the final layer ({\bf Figure \ref{fig:10}}). The numbers of hidden layers and nodes per layer as well as the dropout and learning rates are optimized as hyperparameters over a grid of values based on predictive performance. We implement our approach using the R interface for the deep learning library TensorFlow \citep{tensorflow2015-whitepaper}, with model building and fitting done using Keras API \citep{chollet2015keras}. Finite dimensional parameter training is done via the GradientTape API \citep{agrawal2019tensorflow} for automatic differentiation. Intensive simulations have indicated the new method predicts the risks well (Supplement A).

Revisiting  the BLCSC study, we exemplify our method by studying the impact of clinical and genetic predictors on disease progression and mortality. The subset includes 5,296 patients with non-small cell lung cancer, diagnosed between June 1983 and October 2021. Also included in the dataset are patients' characteristics, namely, age at diagnosis (years), sex (0:~male; 1:~female), race (0:~other; 1:~white), ethnicity (0:~non-Hispanic; 1:~Hispanic), height (meters), weight (kilograms), smoking status (0:~never; 1:~former; 2:~current), pack-years, cancer stage (1-4), and two indicators of genetic mutations (EGFR and KRAS). 
Semi-competing events of cancer progression and death are documented in the data;
the date of progression is the date of the first source evidence, including exam, radiology report or  pathology. Progression followed by death is observed in 111 (2\%) patients, progression but alive at the last followup date is observed in 224 (4\%) patients, and death prior to progression is observed among 1,916 (36\%) patients. 

To investigate the dependence of disease progression on death and predict the transition processes, we fit models (\ref{eq:condhaz1})-(\ref{eq:condhaz3}) via a DNN-SCR. Specifically, we assume Weibull baseline hazards $\lambda_{0g}(s), g=1, 2,3$, and $\gamma_{i} \stackrel{i.i.d}{\sim} \Gamma(1/\theta, 1/\theta)$ [as specified underneath (\ref{eq:condhaz3})], and let $\boldsymbol{X}_i$ be the $i$th patient's characteristics, $i=1, \ldots, 5,296$. We then use DNN-SCR to optimize the objective function (\ref{eq:obj}) in order to output the estimates of the finite dimensional parameters ($\phi$'s and $\theta$) and the predicted $h_g, g=1,2,3$ (log risk estimates), for any covariate values.

\begin{figure}[!ht]
    \centering
    \includegraphics[width = \textwidth]{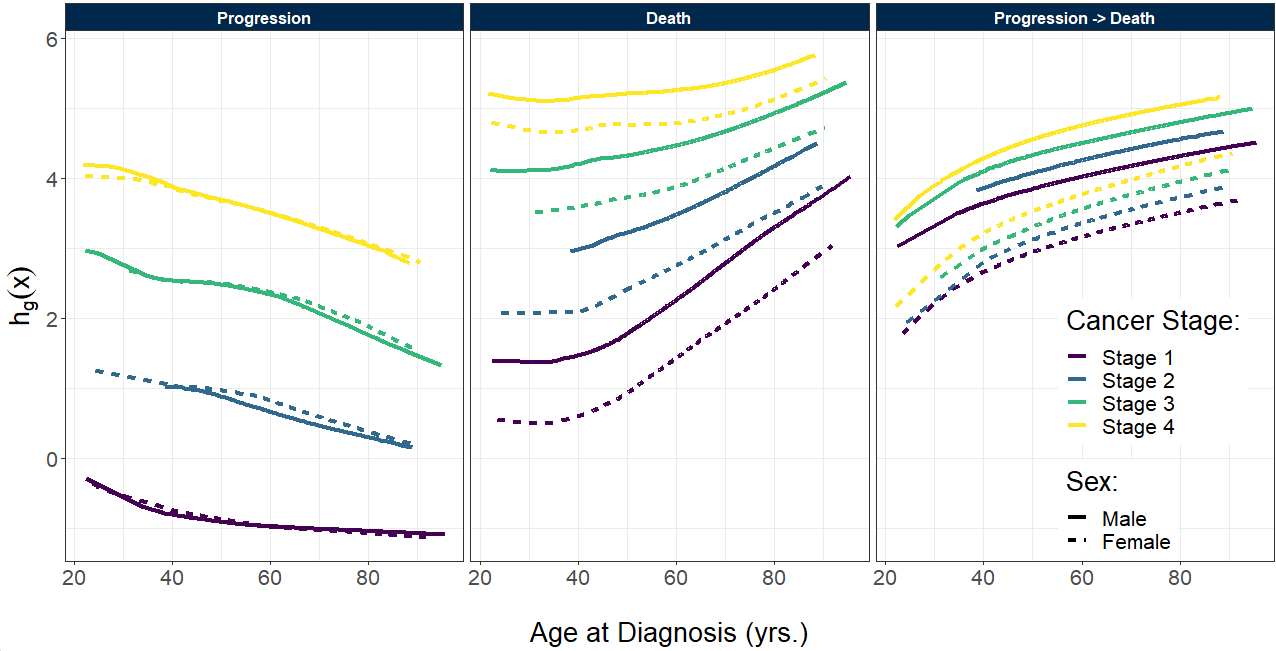}
    \caption{Log risk functions of age at diagnosis on each state transition, stratified by sex (solid versus dashed lines) and initial cancer stage (line color).}
    \label{fig:11}
\end{figure}

We estimate the frailty variance, $\theta$, to be 3.15 (bootstrapped 95\% CI: 3.02-3.29), suggesting that progression is indeed correlated with death. {\bf Figure \ref{fig:11}} depicts the $h$ functions (log risks) for the effect of age at diagnosis on each state transition, stratified by sex and initial cancer stage while fixing the other covariates to be at their
sample means or modes. There seems to exist a non-linear effect of age that differs by transition,   cancer stage and sex. The left panel shows that younger age and more advanced stage is  associated with higher hazards for progression; for the transitions from diagnosis or progression to death (the middle and right panels),  older age is associated with higher hazards; interestingly, while sex does not seem to play a role in disease progression (the left panel), male patients are more likely to die than female patients after diagnosis  (the middle panel) or after progression (the right panel). Finally, more advanced stage is associated with  higher hazards for all the transitions.
 
\section{CONCLUSIONS}

We have presented various methods for analyzing survival outcome data with high-dimensional predictors. We first provided a primer on time to event data and the unique features of survival analysis that make it distinct from other areas of statistics. We then reviewed regularization approaches for extending classical models such as the Cox, AFT, and censored quantile regression models, which lay the foundation for much of the subsequent work in this field, to high-dimensional settings.  We briefly touched on feature screening  for ultra high-dimensional predictors and  discussed high dimensional inference with survival data. Finally, we focused on machine learning for survival prediction and concluded with methods at the forefront of the field of prognostication with competing event data. 

This review is intended to provide a roadmap for readers interested in high-dimensional survival analysis (see Supplement B for tabulation of  the reviewed methods and their available software), though our review is by no means exhaustive.  This is an exciting and rapidly evolving field, with many open questions and new developments. For example, progress in survival predictions with high-dimensional predictors, including deep learning, active learning, and transfer learning will open new avenues to interdisciplinary breakthroughs in biomedical research and data-driven prognostic methods. Also, our review is mainly focused on frequentist methods, and in the last decade, a significant portion of Bayesian works \citep{lee2011bayesian, annest2009iterative, wang2013ibag, pungpapong2021incorporating, bonato2011bayesian} have appeared, which make the field even more exciting. The paper pays tribute to the late Sir D.R. Cox, whose work in survival analysis has fundamentally changed the paradigm of biomedical research and will continue to impact future research for years to come.


\section*{DISCLOSURE STATEMENT}

The authors declare no affiliations, memberships, funding, or financial holdings that might be perceived as affecting the objectivity of this work. 


\section*{ACKNOWLEDGMENTS}

We thank our long-term collaborator, Dr. David C. Christiani, for providing the BLCSC data and Dr. Xinan Wang for helpful discussion of the BLCSC application results. We thank Dr. Ingrid van Keilegom and a reviewer for many helpful suggestions that significantly improved the quality of the paper. The work is partially supported by the grants from NIH.


\bibliography{ref}

\end{document}


\noindent {\large \bf Supplemental Materials}\\

\noindent The following document provides supporting information for the manuscript:~``High-Dimensional Survival Analysis: Methods and Applications.''\\

\pagebreak

\noindent {\bf A. Simulation Results for the Proposed Deep Neural Network for Semi-Competing Risks} \\

\large{
\noindent We conduct simulations to illustrate the feasibility of the proposed Deep Neural Network for Semi-Competing Risks (DNN-SCR) model.\\

We simulate the observed data, $\mathcal{D} = \{(Y_{i1},\ \delta_{i1},\ Y_{i2},\ \delta_{i2},\ x_i);\ i=1,\ldots,n\}$ in a fully factorial design by varying the sample size, frailty variance, log-risk function, and censoring rates, a total of 24 settings (Table \ref{tab:sim}). \\

Specifically, we simulate the shared frailty, $\gamma_i$, from $\Gamma(1/\theta, 1/\theta)$  with $ {\rm Var}(\gamma_i)=\theta$ taking values of 0.5 and 2.0, corresponding to varying degrees of dependence between event times.\\ 

The baseline hazard functions, $\lambda_{01}$, $\lambda_{02}$, and $\lambda_{03}$, are taken to be  Weibull distributions with the same shape and scale parameters equal to 1. We simulate two standard Normal random covariates, $X_1, X_2\sim N(0,1)$, which are taken to be predictive of the morbidity and  mortality hazards through both a linear and non-linear log-risk function. 
Specifically,  we first examine a linear log-risk function: 
$$h_g(\boldsymbol{X}_i) = x_{i}^\top \boldsymbol{\beta}_{g}
$$
with $\boldsymbol{\beta}_{g} = [1, 1]^\top$ for  $g = 1,2,3,$
so that the requirements for the classical model is satisfied, facilitating a fair comparison with existing methods. Then, we consider non-linear functions $$h_g(\boldsymbol{X}_i) = \log(|\boldsymbol{X}_{i}|^\top \boldsymbol{\beta}_{g}+1)$$   with $\boldsymbol{\beta}_{g} = [1, 1]^\top$ for  $g = 1,2,3$.\\
 
Censoring times are generated from an exponential distributions to yield approximate censoring rates of 0\%, 25\% and 50\%. We vary the number of patients as $1,000$ and $10,000$. For each parameter configuration, a total of 50 datasets are independently generated.\\

We compared our method to a classical  MLE approach, which directly maximizes the log-likelihood function under the assumption of a semi-Markov model with Weibull baseline hazard functions. This approach assumes that the risk functions are linear combinations of the generated covariates. We compare the predictive performance of our method to the MLE approach using the average mean integrated squared error for estimating the log-risk surfaces: $$ \frac{1}{n} \sum_{i=1}^n [h_g(\boldsymbol{X}_i) - \hat{h}_g(\boldsymbol{X}_i)]^2; g = 1,2,3,$$ for each state transition hazard, separately. \\

As shown in Table \ref{tab:sim}, both methods accurately recover the log-risk surfaces for each state transition when the true underlying function is linear. However, in the non-linear settings, our deep neural network approach has a much lower mean integrated squared error, on average, compared to the classical MLE method, indicating a good performance of the proposed method.
}
\pagebreak

\setcounter{table}{0}
\renewcommand{\thetable}{A\arabic{table}}

\begin{table}[!ht] 

\centering
\caption{Average (SD) Mean Integrated Squared Errors for Simulated Log-Risk Surfaces for Each State Transition Hazard (i.e., $1/n \sum_{i=1}^n [h_g(x_i) - \hat{h}_g(x_i)]^2; g = 1,2,3$)}
\label{tab:sim}
\begin{longtable}{| p{.075\linewidth} | p{.075\linewidth} | p{.075\linewidth} | p{.075\linewidth} | p{.075\linewidth} | p{.075\linewidth} | p{.075\linewidth} | p{.075\linewidth} | p{.075\linewidth} | p{.075\linewidth} | p{.075\linewidth} |} 
  \hline
\multicolumn{5}{c}{Simulation Settings} & \multicolumn{3}{c}{Maximum Likelihood Estimation} & \multicolumn{3}{c}{Deep Neural Network} \\
\hline
Setting & Sample \hskip 5ex Size & Frailty\hskip 5ex Variance ($\theta$) & Log-Risk Function & Censoring Rate   &       $h_1$ &       $h_2$ &       $h_3$ &       $h_1$ &       $h_2$ &       $h_3$ \\ 
  \hline
      1 &  1,000 & 0.50 & Linear     &  0\% & 0.01 (0.01) & 0.01 (0.01) & 0.01 (0.01) & 0.07 (0.05) & 0.08 (0.08) & 0.08 (0.05) \\ 
      2 & 10,000 & 0.50 & Linear     &  0\% & 0.00 (0.00) & 0.00 (0.00) & 0.00 (0.00) & 0.08 (0.07) & 0.08 (0.05) & 0.08 (0.07) \\ 
      3 &  1,000 & 2.00 & Linear     &  0\% & 0.02 (0.01) & 0.01 (0.01) & 0.01 (0.01) & 0.12 (0.07) & 0.13 (0.07) & 0.13 (0.09) \\ 
      4 & 10,000 & 2.00 & Linear     &  0\% & 0.00 (0.00) & 0.00 (0.00) & 0.00 (0.00) & 0.11 (0.06) & 0.11 (0.08) & 0.13 (0.10) \\ 
      \hline
      5 &  1,000 & 0.50 & Non-Linear &  0\% & 1.80 (0.33) & 1.82 (0.39) & 1.85 (0.34) & 0.09 (0.05) & 0.09 (0.04) & 0.08 (0.04) \\ 
      6 & 10,000 & 0.50 & Non-Linear &  0\% & 1.80 (0.13) & 1.77 (0.13) & 1.78 (0.11) & 0.07 (0.03) & 0.08 (0.03) & 0.08 (0.05) \\ 
      7 &  1,000 & 2.00 & Non-Linear &  0\% & 1.92 (0.53) & 1.85 (0.54) & 1.96 (0.53) & 0.15 (0.05) & 0.15 (0.06) & 0.14 (0.05) \\ 
      8 & 10,000 & 2.00 & Non-Linear &  0\% & 1.82 (0.17) & 1.81 (0.18) & 1.83 (0.18) & 0.14 (0.04) & 0.12 (0.03) & 0.13 (0.06) \\ 
      \hline
      9 &  1,000 & 0.50 & Linear     & 25\% & 0.01 (0.02) & 0.02 (0.01) & 0.02 (0.02) & 0.10 (0.06) & 0.10 (0.07) & 0.13 (0.12) \\ 
     10 & 10,000 & 0.50 & Linear     & 25\% & 0.00 (0.00) & 0.00 (0.00) & 0.00 (0.00) & 0.12 (0.10) & 0.12 (0.09) & 0.12 (0.10) \\ 
     11 &  1,000 & 2.00 & Linear     & 25\% & 0.03 (0.02) & 0.02 (0.02) & 0.04 (0.03) & 0.15 (0.10) & 0.13 (0.09) & 0.18 (0.12) \\ 
     12 & 10,000 & 2.00 & Linear     & 25\% & 0.00 (0.00) & 0.00 (0.00) & 0.00 (0.00) & 0.14 (0.10) & 0.12 (0.08) & 0.14 (0.10) \\ 
     \hline
     13 &  1,000 & 0.50 & Non-Linear & 25\% & 1.96 (0.44) & 2.01 (0.54) & 2.24 (0.66) & 0.10 (0.07) & 0.10 (0.06) & 0.10 (0.08) \\ 
     14 & 10,000 & 0.50 & Non-Linear & 25\% & 1.95 (0.15) & 1.91 (0.16) & 2.16 (0.20) & 0.07 (0.04) & 0.09 (0.08) & 0.09 (0.07) \\ 
     15 &  1,000 & 2.00 & Non-Linear & 25\% & 2.06 (0.62) & 1.92 (0.72) & 2.25 (0.79) & 0.15 (0.08) & 0.15 (0.08) & 0.13 (0.06) \\ 
     16 & 10,000 & 2.00 & Non-Linear & 25\% & 1.88 (0.21) & 1.88 (0.21) & 2.04 (0.28) & 0.10 (0.05) & 0.11 (0.06) & 0.11 (0.05) \\ 
     \hline
     17 &  1,000 & 0.50 & Linear     & 50\% & 0.01 (0.02) & 0.02 (0.02) & 0.04 (0.03) & 0.10 (0.07) & 0.10 (0.06) & 0.20 (0.15) \\ 
     18 & 10,000 & 0.50 & Linear     & 50\% & 0.00 (0.00) & 0.00 (0.00) & 0.00 (0.01) & 0.10 (0.07) & 0.11 (0.08) & 0.17 (0.16) \\ 
     19 &  1,000 & 2.00 & Linear     & 50\% & 0.03 (0.03) & 0.03 (0.02) & 0.05 (0.05) & 0.22 (0.13) & 0.17 (0.13) & 0.24 (0.17) \\ 
     20 & 10,000 & 2.00 & Linear     & 50\% & 0.00 (0.00) & 0.00 (0.00) & 0.01 (0.00) & 0.14 (0.09) & 0.14 (0.10) & 0.16 (0.14) \\ 
     \hline
     21 &  1,000 & 0.50 & Non-Linear & 50\% & 2.06 (0.50) & 2.20 (0.72) & 2.61 (1.00) & 0.09 (0.06) & 0.13 (0.13) & 0.18 (0.14) \\ 
     22 & 10,000 & 0.50 & Non-Linear & 50\% & 2.03 (0.21) & 2.00 (0.18) & 2.36 (0.25) & 0.06 (0.03) & 0.09 (0.08) & 0.10 (0.09) \\
     23 &  1,000 & 2.00 & Non-Linear & 50\% & 2.16 (0.76) & 2.00 (0.72) & 2.41 (0.91) & 0.18 (0.10) & 0.18 (0.09) & 0.16 (0.10) \\ 
     24 & 10,000 & 2.00 & Non-Linear & 50\% & 1.92 (0.25) & 1.95 (0.23) & 2.22 (0.38) & 0.10 (0.05) & 0.11 (0.06) & 0.15 (0.13) \\ 
   \hline
\end{longtable}
\end{table}

\pagebreak

\noindent {\bf B. Selected Methods, Citations, and Available Software}\\

\setcounter{table}{0}
\renewcommand{\thetable}{B\arabic{table}}

\begin{table}[!ht]
    \tabcolsep7.5pt
    \caption{Selected methods covered in this review, citations, and available software}
    \scriptsize
    \begin{longtable}{| p{.31\linewidth} | p{.2\linewidth} | p{.4\linewidth} |} 
    \hline 
        Method & Citation & Available Software\\
        \hline 
        {\bf Classical Survival Analysis} & & \\
        \quad Cox Proportional Hazards Model & \cite{Cox1972}  & https://cran.r-project.org/web/packages/survival/index.html \\
        \quad Accelerated Failure Time Models & \cite{buckley1979linear} & https://cran.r-project.org/web/packages/survival/index.html \\
        \quad Censored Quantile Regression &  \cite{portnoy2003censored} & https://cran.r-project.org/web/packages/quantreg/index.html \\
        \quad Sub-Distribution Hazard Model & \cite{fine1999proportional} & https://cran.r-project.org/web/packages/cmprsk/index.html \\
        \quad Illness-Death Model & \cite{haneuse2016semi} & https://cran.r-project.org/web/packages/SemiCompRisks/index.html \\
        \hline
        {\bf Regularized Cox Models} & & \\
        \quad Ridge           & \cite{verweij1994penalized} & https://cran.r-project.org/web/packages/glmnet/glmnet.pdf \\
        \quad LASSO           & \cite{Tibshirani1997}         & https://cran.r-project.org/web/packages/glmnet/glmnet.pdf \\
        \quad Elastic Net     & \cite{wu2012elastic}          & https://cran.r-project.org/web/packages/glmnet/glmnet.pdf \\
        \quad Adaptive LASSO  & \cite{ZhangLu2007}            & https://cran.r-project.org/web/packages/glmnet/glmnet.pdf \\
        \quad SCAD & \cite{FanLi2002} & https://cran.r-project.org/web/packages/ncvreg/index.html \\
        \quad Group LASSO     & \cite{kim2012analysis}        & https://cran.r-project.org/web/packages/grpreg/index.html \\
        \quad Fused LASSO     & \cite{tibshirani2005sparsity} & https://cran.r-project.org/web/packages/penalized/penalized.pdf \\
        \quad Graphical LASSO & \cite{friedman2008sparse}     & -\\
        \hline
        {\bf Dantzig Selector} & & \\
        \quad AFT Models & \cite{li2014dantzig} & http://www-personal.umich.edu/~yili/adsfxns.R\\
        \quad Cox Model & \cite{antoniadis2010dantzig} & -\\
        \hline
        {\bf Ultra High-Dimensional Feature Screening} & & \\
        \quad Sure Independence Screening & \cite{FanLv2008} & https://cran.r-project.org/web/packages/SIS/index.html\\
        & \cite{FanSong2010} & https://cran.r-project.org/web/packages/SIS/index.html\\
        \quad Principled Sure Independent Screening &  \cite{zhao2012principled} & http://faculty.washington.edu/acook/software.html\\
        \quad Buckley-James Assisted Sure Screening & \cite{liu2020new} & - \\ 
        \quad Conditional Screening & \cite{kang2017partition} & https://github.com/younghhk/software/blob/master/CS.R\\
        \quad Forward Regression & \cite{hong2019quantile} & -\\  
        {\bf Inferential Methods} & & \\
        \quad Selection-Assisted Partial Regression and Smoothing & \cite{fei2021estimation} & https://github.com/feizhe/SPARES\\
        \quad Fused High-Dimensional Censored Quantile Regression & \cite{fei2021inference} & https://github.com/feizhe/HDCQR\_Paper \\
        \hline
        {\bf Support Vector Machines} & & \\
        \quad Rank-Based Approach & \cite{van2007support} & https://cran.r-project.org/web/packages/survivalsvm/index.html \\
        \quad Regression Approach & \cite{shivaswamy2007support} & https://cran.r-project.org/web/packages/survivalsvm/index.html \\
        \quad Hybrid Approach & \cite{van2011support} & https://cran.r-project.org/web/packages/survivalsvm/index.html \\ 
        & \cite{polsterl2015fast} & https://cran.r-project.org/web/packages/survivalsvm/index.html \\
        \hline
        {\bf Tree-Based Methods} & & \\
        \quad Log-Rank Based &  \cite{ciampi1986stratification} & https://cran.r-project.org/web/packages/rpart/index.html\\
        \quad Likelihood-Based & \cite{ciampi1987recursive} & https://cran.r-project.org/web/packages/rpart/index.html\\
        \hline
        {\bf Ensemble Learners} & & \\
        \quad Bootstrap Aggregation & \cite{hothorn2004bagging} & https://cran.r-project.org/web/packages/ipred/index.html \\
        \quad Gradient Boosting & \cite{hothorn2006survival} & https://cran.r-project.org/web/packages/gbm/index.html \\
        \quad Random Survival Forests & \cite{ishwaran2008random} & https://cran.r-project.org/web/packages/randomForestSRC/index.html \\
        & \cite{ishwaran2011random} & https://cran.r-project.org/web/packages/randomForestSRC/index.html \\
        & \cite{ishwaran2019standard} & https://cran.r-project.org/web/packages/randomForestSRC/index.html \\
        \quad Censoring Unbiased Regression Trees & \cite{steingrimsson2019censoring} & https://cran.r-project.org/web/packages/randomForest/index.html \\
        \hline
        {\bf Deep Learning} & & \\
        \quad DeepSurv & \cite{katzman2018deepsurv} & https://cran.r-project.org/web/packages/survivalmodels/index.html \\
        \quad DNNSurv & \cite{zhao2020deep} & https://cran.r-project.org/web/packages/survivalmodels/index.html \\
        \hline
        {\bf Competing Risks} & & \\
        \quad DeepHit & \cite{lee2018deephit} & https://cran.r-project.org/web/packages/survivalmodels/index.html \\
        \quad Dynamic DeepHit & \cite{lee2019dynamic} & https://github.com/chl8856/Dynamic-DeepHit\\
        \quad  DeepCompete & \cite{aastha2020deepcompete} & - \\
        \quad Hierarchical Multi-Event  & \cite{tjandra2021hierarchical} & - \\
        \hline
        {\bf Semi-Competing Risks} & & \\
        \quad Penalized Estimation & \cite{reeder2022penalized} & - \\
        \quad Deep Learning & {\it Proposed} & - \\
        \hline
    \end{longtable}
    \label{tab:methods}
\end{table}

\pagebreak

\small

\bibliography{ref}